\documentclass[twocolumn,superscriptaddress,showpacs,preprintnumbers,amsmath
,amssymb,aps,prd]{revtex4}
\usepackage{graphicx}
\usepackage{bm}
\usepackage{color}
\usepackage{hyperref}
\usepackage{verbatim}
\usepackage[caption=false]{subfig}
\usepackage{units}

\def\bbh#1{binary black hole#1 (BBH#1)\gdef\bbh{BBH}}
\def\bh#1{black hole#1 (BH#1)\gdef\bh{BH}}
\def\ns#1{neutron star#1 (NS#1)\gdef\ns{NS}}
\def\hmns#1{hypermassive neutron star#1 (HMNS#1)\gdef\hmns{HMNS}}
\def\qnm#1{quasi-normal-mode#1 (QNM#1)\gdef\qnm{QNM}}
\def\bhns#1{black hole - neutron star#1 (BH-NS#1)\gdef\bhns{BH-NS}}
\def\dns#1{double neutron star#1 (DNS#1)\gdef\dns{DNS}}
\def\pnw#1{post-Newtonian#1 (PN#1)\gdef\pnw{PN}}
\def\gw#1{gravitational wave#1 (GW#1)\gdef\gw{GW}}
\def\W{{\cal W}}  

\def\newacronym#1#2#3{\gdef#1{#3 (#2)\gdef#1{#2}}}

\newacronym{\ligo}{LIGO}{Laser Interferometer Gravitational Wave Observatory}
\newacronym{\tov}{TOV}{Tolman-Oppenheimer-Volkoff}
\newacronym{\nr}{NR}{numerical relativity}

\newcommand{\carpet}{\textsf{Carpet}}
\newcommand{\cactus}{\textsf{Cactus}}
\newcommand{\kranc}{\textsf{Kranc}}
\newcommand{\maya}{\textsf{Maya}}
\newcommand{\twopunctures}{\textsf{2Punctures}}
\newcommand{\lorene}{\textsf{Lorene}}
\newcommand{\ET}{\textsf{Einstein Toolkit}}
   
\begin{document}

\title{Bowen-York Type Initial Data for Binaries with Neutron Stars}

\author{Michael Clark}
\affiliation{Center for Relativistic Astrophysics and
School of Physics\\
Georgia Institute of Technology, Atlanta, GA 30332}
\author{Pablo Laguna}
\affiliation{Center for Relativistic Astrophysics and
School of Physics\\
Georgia Institute of Technology, Atlanta, GA 30332}

\begin{abstract} 
A new approach to construct initial data for binary systems with neutron star components is introduced. The approach is a generalization of the puncture initial data method for binary black holes based on Bowen-York solutions to the momentum constraint. As with binary black holes, the method allows setting orbital configurations with direct input from post-Newtonian approximations and involves solving only the Hamiltonian constraint. The effectiveness of the method is demonstrated with evolutions of double neutron star and black hole -- neutron star binaries in quasi-circular orbits.
  
\end{abstract}

\maketitle


\section{Introduction}
\label{sec:intro}

Compact object binaries with \bh{} and \ns{} components are main targets of \gw{} observations. \gw{s} from \bbh{s} have been recently detected by the \ligo{,} first detection in the transient event GW150914~\cite{2016PhRvL.116f1102A} and second detection in the transient event GW151226~\cite{PhysRevLett.116.241103}. As advanced \ligo{} reaches designed sensitivity, \gw{s} from  \dns{} and \bhns{} binaries will very likely also be detected. Not surprisingly, \nr{} simulations played an important role in the analysis of the GW150914 and GW151226 events. Specifically, best fits of a \nr{} waveform to the data were included in the  detection paper~\cite{2016PhRvL.116f1102A}. The papers on parameter estimation~\cite{2016arXiv160203840T} and tests of general relativity~\cite{2016arXiv160203841T} mentioned that results from \bbh{} simulations were involved in the construction of the phenomenological and effective-one-body waveform models used in the analysis. The same applies to the paper on the burst-type analysis of GW150914~\cite{2016arXiv160203843T}.

As with GW150914 and GW151226, our ability to distinguish in future \gw{} observations whether a signal originated from a \bbh{}, a \dns{}, or a \bhns{} binary will rely on waveform templates with input from \nr{.}  This would be particularly important during the last orbits and coalescence of the binary, where strong dynamical gravity is the most relevant.  In this regard, \nr{} simulations of binary systems with \ns{} companions have experience a boost in accuracy and sophistication. These days the simulations routinely include realistic equations of state, magnetic fields, and radiation. But the predicting power of simulations  not only hinges on the multi-physics included. The degree to which the initial data represent an accurate astrophysical setting is also crucial. Another important aspect connected to the initial data is the capability to explore a vast range of scenarios. And for this to happen, one needs initial data methodologies that are flexible and computationally inexpensive. In \bbh{} simulations, low-cost and efficient methods to construct astrophysically relevant initial data have been available for some time~\cite{Brandt:1997tf,Ansorg:2004ds,Tichy:2003zg}, which is not exactly the case for binaries with \ns{s}. 

A popular method to construct initial data representing a binary system in a quasi-circular orbit is the conformal thin sandwich approach.  The method has been used for \bbh{s} by~\citet{grandclement-2002}, for \dns{} by~\citet{gourgoulhon-2001}, and for \bhns{} binaries by~\citet{etienne-2008}.  The key in those studies was the identification of a helical Killing vector field, so the initial data are approximately time-symmetric, ensuring that the compact objects are in a quasi-circular orbit. The conformal thin sandwich approach requires solving a set of five elliptic equations for the conformal factor, lapse function and shift vector ~\cite{2010nure.book.....B,2011LRR....14....6S,2012LRR....15....8F}. Many groups have used the \lorene{} code from the Meudon group~\cite{2010CQGra..27k4105R,2013PhRvD..87f4023R} for this purpose, and other groups have developed their own infrastructure~\cite{2012PhRvD..86f4024T,2009CQGra..26q5018T,2013PhRvD..88f4060T,2007PhRvD..75h4005T,2013PhRvD..87h4006F,PhysRevD.77.124051,2009PhRvD..79l4018K}.

This paper introduces a new approach to construct initial data for binary systems with \ns{} components. The method is simpler than the thin sandwich one, and it has a computational cost and flexibility similar to that of the \bbh{} puncture method. In the \bbh{} puncture approach~\cite{PhysRevD.77.044037}, one only solves the Hamiltonian constraint for the conformal factor. The solution to the momentum constraint is given by the Bowen-York extrinsic curvature~\cite{1980PhRvD..21.2047B}. Each initial data set is then fully specified by the masses, spins and momenta of the \bh{s}, and their separation. All of these parameters are obtained from integrating the \pnw{} equations of motion. The integration starts at large separations and ends at the separation where the \nr{} initial data are constructed. This method is known to yield initial data suitable for stitching together \nr{} and \pnw{} evolutions.

The new initial data proposal in this paper recycles most of the elements of the puncture  \bbh{} initial data.  The key step is constructing an extrinsic curvature for \ns{s} similar to the Bowen-York for \bh{s}. The paper is organized as follows: In Section~\ref{sec:initial}, we provide a quick review of York's initial data formulation. Section~\ref{sec:K} reintroduces the Bowen-York extrinsic curvature for arbitrary, spherically symmetric momentum sources. Section~\ref{sec:sources} discusses an approach to specifying the matter source functions for the initial data equations. Section~\ref{sec:procedure} summarizes the steps to construct initial data. Section~\ref{sec:TOV} reviews the stellar model we will use to represent \ns{s}. Section~\ref{sec:single} presents tests with an isolated \ns{}. Results of simulations of \dns{} and \bhns{} binaries are presented in Section~\ref{sec:results}. Paper ends with conclusions in Section~\ref{sec:end}.

The numerical simulations in the present work were carried out with our \maya{} code~\cite{Haas:2012bk,2012CQGra..29w2002H,Bode:2011xz,Bode:2011tq,Bode:2009mt,Healy:2009zm}. The code is based on the BSSN formulation of the Einstein equations~\cite{Baumgarte99} and the moving puncture gauge condition~\cite{Campanelli:2005dd,Baker:2005vv}.  \maya{}  is very similar to the Einstein code in the \ET{} \cite{et-web}. That is, it operates under the  \cactus{} infrastructure~\cite{Allen99a}, with \carpet{} providing mesh refinements~\cite{Schnetter-etal-03b} and thorns (modules) generated by
the package \kranc{}~\cite{Husa:2004ip}. 

\section{Initial Data at a Glance}
\label{sec:initial}

When the Einstein equations of general relativity are viewed as an initial value problem, the initial data are not completely freely specifiable. They must satisfy the Hamiltonian and momentum constraints:
\begin{eqnarray}
R + K^2 - K_{ij}K^{ij} &=& 16\,\pi\,\rho_{\rm H}  \label{eq:Ham}\\
\nabla_j (K^{ij} - \gamma^{ij}K) &=& 8\,\pi\,S^i \label{eq:Mom}\,. 
\end{eqnarray}
Above, $\gamma_{ij}$ and $K_{ij}$ are the metric and extrinsic curvature of the space-like hypersurfaces in the foliation. In addition, $R$ is the Ricci scalar, and $\nabla_i$ denotes covariant differentiation associated with  $\gamma_{ij}$. The sources $\rho_{\rm H}$ and $S^i$ are obtained from the stress-energy tensor $T_{ab}$ as follows:
\begin{eqnarray}
\rho_{\rm H} &=& n^an^b T_{ab}\label{eq:rho}\\
S^i &=& -\gamma^{ib}n^cT_{bc}\label{eq:si} \,,
\end{eqnarray}
where $n^a$ is the unit normal to the space-like hypersurfaces. We are using units in which $G=c=1$. Latin indices from the beginning of the alphabet denote spacetime indices and from the middle of the alphabet spatial indices. For a perfect fluid, the stress-energy tensor reads
\begin{eqnarray}
T_{ab} &=& (\rho+p)\,u_au_b + p\,g_{ab}\nonumber\\
            &=&\rho_0\,h\,u_au_b + p\,g_{ab}\label{eq:Tab}\,,
\end{eqnarray}
where $h= 1+\epsilon+p/\rho_0$ is the enthalpy, $p$ is the pressure, $u^a$ is the 4-velocity of the fluid, 
$\rho_0$ is the rest-mass density, $\epsilon$ is the specific internal energy density, and $\rho = \rho_0(1+\epsilon)$ is the total mass-energy density. In terms of these quantities, the sources in the Hamiltonian and momentum constraints read:
\begin{eqnarray}
\rho_{\rm H} &=& (\rho+p)\,W^2 - p = \rho_0\,h\,W^2 - p\label{eq:rho_fluid}\\ 
S^i &=&  (\rho + p) W u^i = \rho_0\,h\, W u^i\label{eq:si_fluid} 
\end{eqnarray}
where $W = -n_au^a$ is the Lorentz factor between normal and fluid observers. 

Since the initial data consist of the set $\{ \gamma_{ij}, K_{ij}, \rho_{\rm H}, S^i \}$, the pressing issue is to identify which ``pieces'' in these data are to be fixed by the constraint Eqs.~ (\ref{eq:Ham}) and (\ref{eq:Mom}), and which data are indeed freely specifiable. 

Motivated by the work of Lichnerowicz~\cite{2010nure.book.....B}, York and collaborators~\cite{1979sgrr.work.....S} developed an elegant way of achieving this task. The basis of this approach is using conformal transformations and transverse-traceless decompositions to single out the four quantities fixed by the constraint equations. The transformations and decompositions are:
\begin{eqnarray}
\gamma_{ij} &=& \Phi^4\bar\gamma_{ij}\\
K_{ij} &=& A_{ij} + \frac{1}{3}\gamma_{ij} K\\
A_{ij} &=& \Phi^{-2}\bar A_{ij}\\
\bar A_{ij} &=& \bar A^{\rm TT}_{ij} + \bar A^{\rm L}_{ij}\,.
\end{eqnarray}
With them,  Eqs.~(\ref{eq:Ham}) and (\ref{eq:Mom}) reduce to
\begin{eqnarray}
8\bar\Delta \Phi - \Phi\,\bar R - \frac{2}{3} \Phi^5 K^2&+& \Phi^{-7} \bar A_{ij} \bar A^{ij}  \nonumber \\
&=& - 16\,\pi \Phi^5\,\rho_{\rm H} \label{eq:HamC}\\
(\bar\Delta_{\rm L} \W)^i - \frac{2}{3}\Phi^6\, \bar\nabla^i K &=& 8\,\pi\,\Phi^{10}S^i\label{eq:MomC}
\end{eqnarray}
respectively, with 
\begin{eqnarray}
\bar A_{\rm L}^{ij} &=& (\bar L \W)^{ij} \\
\bar\nabla_i \bar A^{ij}_{\rm TT} &=& 0\\
 (\bar L \W)^{ij} &\equiv& \bar\nabla^i\W^j + \bar\nabla^j \W^i 
- \frac{2}{3} \bar\gamma^{ij} \bar\nabla_k\W^k\\
 (\bar\Delta_L \W)^i &\equiv&  \bar\nabla_j(\bar L \W)^{ij} \,.
\end{eqnarray}

Given Eqs.~(\ref{eq:HamC}) and (\ref{eq:MomC}), constructing initial data translates into specifying the quantities  $\{ \hat\gamma_{ij}, K, \bar A^{\rm TT}_{ij}, \rho_{\rm H}, S^i \}$, and
solving for the conformal factor $\Phi$ and vector $\W^i$. 
A common choice, which we adopt, is to assume conformal flatness ($\bar\gamma_{ij} = \eta_{ij}$), maximal slicing ($K=0$), and $\bar A^{\rm TT}_{ij} = 0$. Under these assumptions, the constraints (\ref{eq:HamC}) and \eqref{eq:MomC} assume the form
\begin{eqnarray}
&&\bar\Delta \Phi
+ \frac{1}{8}\Phi^{-7} \bar A_{ij} \bar A^{ij} = - 2\,\pi \Phi^5\,\rho_{\rm H} \label{eq:HamCF}\\
&&(\bar\Delta_L \W)^i = 8\,\pi\,\Phi^{10}S^i\label{eq:MomCF}
\end{eqnarray}
with $\bar A^{ij} = \bar A_{\rm L}^{ij} =  (\bar L \W)^{ij}$. 
We exploit the freedom to conformally transform $\rho_{\rm H}$ and $S^i$ and set
\begin{eqnarray}
\bar \rho_{\rm H} &=& \rho_{\rm H}\,\Phi^{8} \label{eq:Trho}\,,\\
\bar S^i &=& S^i \,\Phi^{10} \,,\label{eq:Srho}
\end{eqnarray}
and thus Eqs.~(\ref{eq:HamCF}) and (\ref{eq:MomCF}) read
\begin{eqnarray}
&&\bar\Delta \Phi
+ \frac{1}{8}\Phi^{-7} \bar A_{ij} \bar A^{ij} = - 2\,\pi \Phi^{-3}\,\bar\rho_{\rm H} \label{eq:HamCF2}\\
&&(\bar\Delta_L \W)^i = 8\,\pi\,\bar S^i\label{eq:MomCF2}\,.
\end{eqnarray}
The transformations (\ref{eq:Trho}) and (\ref{eq:Srho}), and the expressions (\ref{eq:rho_fluid}) and (\ref{eq:si_fluid}) suggest setting in the stress-energy tensor
$\bar\rho = \Phi^{8}\rho$, $\bar p = \Phi^{8}\,p$ and $\bar u^i = \Phi^2u^i$, and therefore
\begin{eqnarray}
\bar\rho_{\rm H} &=& (\bar\rho+\bar p)\,W^2 - \bar p\,,\label{eq:rhoH}\\
\bar S^i &=& (\bar\rho+\bar p) W\,\bar u^i \,,\label{eq:si2}
\end{eqnarray}
Notice from $u^au_a = -1$ that $W^2 -1 = \gamma_{ij} u^iu^j = \bar \gamma_{ij}\bar u^i\bar u^j = \bar W^2-1$. Then, with the help of Eq.~(\ref{eq:si2}),
\begin{equation}
W^2 -1 =\bar \gamma_{ij}\bar u^i\bar u^j = \frac{\bar S^2}{W^2(\bar\rho + \bar p)^2}\,,
\end{equation}
and thus
\begin{equation}
	W^2 = \frac{1}{2}\left(1 + \sqrt{1 + \frac{4\,\bar S^2}{(\bar\rho + \bar p)^2}}\right) \, ,\label{eq:Lorentz}
\end{equation}
where $\bar S^2 = \bar \gamma_{ij}\bar S^i\bar S^j$.

In summary, constructing initial data reduces to first specifying $\bar\rho_{\rm H}$ and $\bar S^i$, next solving Eq.~(\ref{eq:MomCF2}) for $\W^i$ to construct $\bar A^{ij}$, and finally solving for $\Phi$ from Eq.~(\ref{eq:HamCF2}).

\section{Extrinsic Curvature}
\label{sec:K}

We now consider solutions to the momentum constraint equation $(\bar\Delta_L \W)^i = 8\,\pi\,\bar S^i$. We will first recall 
the solution that represents \bh{s} and next reintroduce the one suitable to model \ns{s}. For \bh{s} ($\bar S^i = 0$), Bowen and York ~\cite{1980PhRvD..21.2047B} found that point-source solutions to $(\bar\Delta_L \W)^i =0$ are given by
\begin{eqnarray}
  \W^i &=& -\frac{1}{4\,r}\left[7\,P^i + l^i(P\cdot l)\right]\label{eq:WP}\\
  \W^i &=& \frac{1}{r^2}\epsilon^{ijk}l_jJ_k\,,\label{eq:WJ}
\end{eqnarray} 
with $l^i = x^i/r$ a unit radial vector and $P\cdot l = P^i l_i$. In these solutions, the constant vectors $P^i$ and $J_i$ are respectively interpreted  as the linear and angular momentum of the \bh{.} From $\bar A^{ij} = (\bar L \W)^{ij}$, the extrinsic curvature associated with these solutions are: 
\begin{eqnarray}
  \bar A^{ij} &=& \frac{3}{2\,r^2}\left[ P^i l^j + P^jl^i - ( \eta^{ij}-l^il^j) (P\cdot l)\right]\label{eq:KP}\\
  \bar A^{ij} &=& \frac{6}{r^3}l^{(i}\epsilon^{j)kl}J_kl_l  \label{eq:KS} 
\end{eqnarray}

Next is to consider solutions to $(\bar\Delta_L \W)^i = 8\,\pi\,\bar S^i$ that can be used to build the extrinsic curvature of a \ns{}. Following Bowen~\cite{1979GReGr..11..227B}, we assume sources of the form
\begin{eqnarray}
\bar S^i &=& P^i\,\sigma(r)\\
\bar S_i &=& \epsilon_{ijk}\,J^jx^k\,\kappa(r)\,.
\end{eqnarray}
At this point, $P^i$ and $J^i$ arbitrary constant vectors, and $\sigma$ and $\kappa$ radial functions with compact support on $r \le r_0$. The specific form of these functions will be determined in the next section using the following conditions.
 
From the definition of ADM linear momentum~\cite{lrr-2011-6}, one has that
\begin{eqnarray}
	P^i _{\rm ADM}&=& \frac{1}{8\pi}\int_{\partial\Sigma_\infty}A^{ij} \,dS_j \nonumber\\
	&=&\frac{1}{8\pi} \int_\Sigma \bar\nabla_j \bar A^{ij} \sqrt{\eta} \,d^3x \nonumber \\
	&=& \int_\Sigma\bar S^{i} \sqrt{\eta}\, d^3x \nonumber \\
	&=&P^i \int_\Sigma\sigma \sqrt{\eta}\, d^3x \,.
\end{eqnarray}
Thus, for $P^i_{\rm ADM} = P^i$ to hold, $\sigma$ must satisfy the following normalization condition:
\begin{equation}
\int_\Sigma\sigma \sqrt{\eta}\, d^3x= 4\,\pi \int_{0}^{r_0} \sigma \,r^2\, dr = 1\,.\label{eq:SigNorm}
\end{equation}

Similarly, from the definition of ADM angular momentum~\cite{2010nure.book.....B}, we have that
\begin{eqnarray}
	J^{\rm ADM}_i &=&  \frac{1}{8\pi}\epsilon_{ijk}\int_{\partial\Sigma_\infty}x^jA^{km} \,dS_m \nonumber\\
	&=&\frac{1}{8\pi}\epsilon_{ijk} \int_\Sigma x^j \bar\nabla_m \bar A^{km} \sqrt{\eta} \,d^3x \nonumber \\
	&=&\epsilon_{ijk} \int_\Sigma x^j  \bar S^{k} \sqrt{\eta} \,d^3x \nonumber \\
         &=&\epsilon_{ijk}\epsilon^{klm} \int_\Sigma x^j  J_l x_m\,\kappa \sqrt{\eta}\,d^3x \nonumber \\
         &=& \int_\Sigma  r^2 \left(J_i - l_i  l^j J_j \right )\kappa \sqrt{\eta} \,d^3x \,.
\end{eqnarray}
Adopting Cartesian coordinates and aligning the angular momentum with the $z$-axis, one gets that
\begin{eqnarray}
	J^{\rm ADM}_i &=& J_i \int_\Sigma r^2\,\sin^2\theta\,\kappa \sqrt{\eta} \,d^3x \,.
\end{eqnarray}
Thus, in order to have $J^{\rm ADM}_i = J_i$,  the following normalization condition must hold
\begin{equation}
2\,\pi \int_0^{r_0}\int_0^{\pi}  \sin^3\theta\,r^4\kappa\,d\theta  \,dr =  \frac{8\,\pi}{3} \int_{0}^{r_0} \kappa \,r^4\, dr = 1\,.\label{eq:KappaNorm}
\end{equation}

Given the normalization condition Eq.~(\ref{eq:SigNorm}) for $\sigma$, the solution to $(\bar\Delta_L \W)^i = 8\,\pi\, P^i\,\sigma$ reads~\cite{1979GReGr..11..227B}
\begin{equation}
  \W^i = -2\,P^iF + \frac{1}{2}P^iH +
  \frac{1}{2} l^i (P\cdot l)\,r H' \label{eq:w}\,.
\end{equation} 
The functions $F$ and $H$ are given respectively by
\begin{eqnarray}
F &=& \frac{1}{r}\int^r_0  \,4\,\pi\,\sigma\,r'^2\,dr'+ \int^{r_0}_r 4\,\pi\,\sigma\,r'\,dr'\label{eq:f}\,,\\
H &=& \frac{1}{r^3}\int^r_0 F\,r'^2\,dr'\hspace{0.5in}\label{eq:h}\,.
\end{eqnarray}

With the help of  $\bar \nabla^i r= l^i$ and $\bar \nabla^i l^j = (\eta^{ij}-l^il^j)/r$, substitution of Eq.~(\ref{eq:w}) into $\bar A^{ij} = (\bar L \W)^{ij}$ yields
\begin{eqnarray}
\label{eq:Kbowen}
\bar A^{ij}&=& (-2\, F' + H')(P^i l^j + P^jl^i)\nonumber \\
&+& (r H''-H') (P \cdot l) l^il^j \nonumber \\
&+&\frac{1}{3}(4\,F' - rH'' - H') (P\cdot l)\eta^{ij}\,.
\end{eqnarray}

With the help of 
\begin{align}
	\label{eq:qjcdefs}
	Q &= \int_0^r 4\pi \sigma r'^2 \, dr' \\
	J &= \int_r^{r_0} 4\pi \sigma r' \, dr' \\
	C &= \int_0^r \frac{2}{3} \pi \sigma r'^4 \, dr'\nonumber\\
	&= \int_0^r \frac{1}{2} Q' r'^2 + \frac{1}{3} J' r'^3 \, dr' \,,
\end{align}
and
\begin{align}
	\label{eq:qjcdefs2}
	F &= Q/r + J \\
	H &= Q/2r + J/3 - C/r^3 \\
	F' &= -Q/r^2 \\
	H' &= -Q/2r^2 + 3C/r^4 \\
	H'' &= Q/r^3 - 12C/r^5 \,,
\end{align}
the expression (\ref{eq:Kbowen})  for the extrinsic curvature can be rewritten as
\begin{eqnarray}
\label{eq:Kbowen2}
\bar A^{ij}&=& \frac{3Q}{2\,r^2}\left[ P^i l^j + P^jl^i - ( \eta^{ij}-l^il^j) (P\cdot l)\right] \nonumber\\
&+& \frac{3 C}{r^4}\left[ P^i l^j + P^jl^i + ( \eta^{ij}-5\,l^il^j) (P\cdot l)\right] \label{eq:KnsP}\,.
\end{eqnarray}

For $r > r_0$ (exterior solution), $Q = 1$, thus the first term in Eq.~(\ref{eq:Kbowen2}) becomes the Bowen-York curvature for a point mass~(\ref{eq:KP}).  Furthermore, Eq.~(\ref{eq:Kbowen2}) has the correct point mass limit since $Q=1$ and $C=0$ for $r_0 = 0$.

For a spherically symmetric source function $\kappa$ with angular momentum $J^i$, the solution to 
$(\bar\Delta_L \W)_i = 8\,\pi\, \epsilon_{ijk} J^jx^k\,\kappa$ is given by~\cite{1989PThPh..81..360O}
\begin{equation}
\W_i = \epsilon_{ijk} \,x^jJ^k\,G \label{eq:wJ}
\end{equation}
where
\begin{eqnarray}
G &=& \frac{1}{r^3}\int^r_0 \frac{8\,\pi}{3}\,r'^4\,\kappa\,dr'+ \int^{r_0}_r\frac{8\,\pi}{3}\kappa\,r'\,dr'\label{eq:g}\,.
\end{eqnarray}
Notice that $G = r^{-3}$ for $r \ge r_0$. Substitution of Eq.~(\ref{eq:wJ}) into $\bar A^{ij} = (\bar L \W)^{ij}$ yields
\begin{eqnarray}
\label{eq:KbowenJ}
 \bar A^{ij} &=& \frac{6}{r^3}l^{(i}\epsilon^{j)kl}J_kl_l N\label{eq:KnsJ}
\end{eqnarray}
where
\begin{equation}
N = \int^r_0 \frac{8\,\pi}{3}\,r'^4\,\kappa\,dr'
\end{equation}
Exterior to the source, $N=1$, and the extrinsic curvature reduces to the point-like solution~(\ref{eq:KS}).
 
In summary, Eqs.~(\ref{eq:KP}) and (\ref{eq:KS}) are the extrinsic curvatures for a point-like source with linear and angular momentum, respectively. In addition,  Eqs.~(\ref{eq:KnsP}) and (\ref{eq:KnsJ}) are the extrinsic curvatures for a spherically symmetric source with linear and angular momentum, respectively. To construct initial data for compact object binaries, the extrinsic curvature for the binary system will be simply given by a superposition of these solutions, point-like for the \bh{} and spherically symmetric source for the \ns{.} The only input needed are the locations of the compact objects, their linear and angular momenta, and the source functions $\sigma$ and $\kappa$.  As with \bbh{s}, the linear and angular momenta of the sources, and their binary separation will be provided by the outcome of integrating the \pnw{} equations of motion. It is very important to keep in mind that, because of the spherical symmetry assumption in the source functions $\sigma$ and $\kappa$, the extrinsic curvature will not be able to account for tidal deformations of the star. We are currently considering a generalization that relaxes the spherical symmetry assumption.

\section{Source Functions}
\label{sec:sources}

The next step is to specify the source functions  $\sigma$ and $\kappa$, as well as the source $\bar\rho_{\rm H} = (\bar\rho+\bar p)\,W^2 - \bar p$ in the Hamiltonian constraint. 
The starting point is the density $\bar \rho$ and  pressure $\bar p$ from the stellar model of our choice,
Recall from Eq.~(\ref{eq:si}) that
$
\bar S^i = (\bar\rho+\bar p) W\,\bar u^i \,.
$
Thus, for the case of linear momentum, we have that 
\begin{equation}
\bar S^i = (\bar\rho+ \bar p)W\, \bar u^i= P^i\,\sigma\,.
\label{eq:p2}
\end{equation}
We then set
\begin{equation}
\sigma  = (\bar\rho+\bar p)/\mathcal M\,, \label{eq:defSig}
\end{equation}
with $\mathcal M$ a constant determined by the normalization condition Eq.~(\ref{eq:SigNorm}) for $\sigma$. That is, 
\begin{equation}
1 = 4\,\pi \int_{0}^{r_0} \sigma \,r^2\, dr = \frac{4\,\pi}{\mathcal M} \int_{0}^{r_0} (\bar\rho+\bar p)\,r^2\, dr \,,
\end{equation} 
and thus
\begin{equation}
{\mathcal M} =  4\,\pi \int_{0}^{r_0} (\bar\rho+\bar p)\,r^2\, dr \,,
\end{equation} 
Notice that Eq.~(\ref{eq:defSig}) restricts our choice for $\bar\rho$ and $\bar p$ to be spherically symmetry solutions since by assumption $\sigma(r)$.
With this choice for $\sigma$, the linear momentum satisfies $P^i = W\,\mathcal M\,\bar u^i$. Since by construction $P^i$ and $\mathcal M$ are constants, $W\,\bar u^i$ must also be constant within the source distribution. 
Finally, notice also from Eqs.~(\ref{eq:Lorentz}), (\ref{eq:p2}) and (\ref{eq:defSig})  that the Lorentz factor is then given by
\begin{equation}
	W^2 = \frac{1}{2}\left(1 + \sqrt{1 + \frac{4\,P^2}{\mathcal M^2}}\right) \, .\label{eq:boostP}
\end{equation}
where $P^2 = \eta_{ij} P^iP^j$. 

For a source with angular momentum, 
\begin{equation}
\bar S_i = \epsilon_{ijk}\,J^jx^k\,\kappa = (\bar\rho+ \bar p)W\, \bar u_i\,.
\end{equation}
As with the previous case, we set
\begin{equation}
\kappa  = (\bar\rho+\bar p)/\mathcal N\,, \label{eq:defKappa}\,.
\end{equation}
From the normalization condition Eq.~(\ref{eq:KappaNorm}), one has that
\begin{equation}
1=  \frac{8\,\pi}{3} \int_{0}^{r_0} \kappa \,r^4\, dr = \frac{8\,\pi}{3\,\mathcal N} \int_{0}^{r_0} (\bar\rho+\bar p) \,r^4\, dr\,,
\end{equation}
and thus the constant $\mathcal N$ is given by
\begin{equation}
{\mathcal N}=   \frac{8\,\pi}{3} \int_{0}^{r_0} (\bar\rho+\bar p) \,r^4\, dr\,,
\end{equation}
The Lorentz factor in this case reads
\begin{equation}
\label{eq:boostJ}
	W^2 = \frac{1}{2}\left(1 + \sqrt{1 + \frac{4\,J^2r^2\sin^2\theta}{\mathcal N^2}}\right) \, .
\end{equation}
where $J$ is the magnitude of the angular momentum and $\theta$ the angle between $J^i$ and $l^i$. It is important to notice that 
in this case the Lorentz boost factor is not constant within the star.

\section{Initial Data Procedure}
\label{sec:procedure}

The centerpiece of our method is solving Eq.~(\ref{eq:HamCF2}), or equivalently
\begin{equation} 
\bar\Delta \Phi + \frac{1}{8}\Phi^{-7} \bar A_{ij} \bar A^{ij} = - 2\,\pi \Phi^{-3}[(\bar\rho+\bar p)\,W^2 - \bar p]    \label{eq:HamCF3}     \,.   
\end{equation}
In this equation, the boost factor $W$ for the stellar model is given by Eq.~(\ref{eq:boostP}) for linear momentum or Eq.~(\ref{eq:boostJ}) for angular momentum. In the same equation, $\bar A^{ij}$ is given by the Bowen-York extrinsic curvatures. For point masses, Eq.~(\ref{eq:KP}) provides the extrinsic curvature with linear momentum and Eq.~(\ref{eq:KS}) the corresponding extrinsic curvature with angular momentum. 
Similarly, the extrinsic curvature associated with the stellar model is given by  Eq.~(\ref{eq:KnsP}) for linear momentum and Eq.~(\ref{eq:KnsJ}) for angular momentum.

In general terms, the sequence of steps to construct initial data for binaries with \bh{s} and \ns{s} components under the proposed method is as follows:

\begin{enumerate}
	\item Choose masses $M_{1,2}$ of the compact objects and their initial separation $d_0$ deep in the \pnw{} regime, with $M = M_1+M_2$ the total mass of the binary and $q = M_1/M_2$ its mass ratio.  Integrate the \pnw{} equations of motion at the highest order available and stop at a separation $d$ where the \nr{} evolution will begin. Read off the linear momentum $\vec P_{1,2}$ and spin $\vec S_{1,2}$ for each of the binary components.
	
	\item Identify the mass $M_{1(2)}$ with the ADM mass $M_{1(2)}^\text{ADM}$ of a star in isolation if a \ns{} and with the irreducible mass $M_{1(2)}^\text{irr}$ if a \bh{}, where
	 \begin{equation}
         \label{eq:madm}
	      M_\text{ADM} = -\frac{1}{2\pi} \int_{\partial\Sigma_\infty} \bar\nabla^i \Phi  \, dS_i\,,
         \end{equation}
	and $M_\text{irr} \equiv \sqrt{{\cal A}/16\,\pi}$ for a \bh{} with apparent horizon area ${\cal A}$~\cite{2010nure.book.....B}.
	
	\item If object 1(2) is a \bh{,} set its puncture bare mass $m_{1(2)} = M_{1(2)}$. If object 1(2) is a \ns{,} construct a spherically symmetric stellar model with ADM mass $M_{1(2)}^\text{ADM}$. Compute also its rest mass $M^0_{1(2)}$ from
          \begin{equation}
	M_0 = \int_{\Sigma} \rho_0 \,W\,\sqrt{\gamma}\, d^3x\,,
          \end{equation}
          and save the ratio $\xi_{1(2)} \equiv M_{1(2)}^\text{ADM}/M^0_{1(2)}$.
	
	\item If the compact object is a \ns{}, calculate the functions $\sigma$ and $\kappa$ from Eqs.~(\ref{eq:defSig}) and (\ref{eq:defKappa}), respectively. 
	
	\item Use the $\vec P$ and $\vec S$ vectors  to construct the extrinsic curvature  using Eqs.~(\ref{eq:KP}) and (\ref{eq:KS}) if a \bh{,} and Eqs.~(\ref{eq:KnsP}) and (\ref{eq:KnsJ}) if a \ns{.} The functions $\sigma$ and $\kappa$ will also be needed if a \ns{}. The total extrinsic curvature is $\bar A^{ij} = \bar A^{ij}_1 +\bar A^{ij}_2$.
	
	\item Construct the term $[(\bar\rho+\bar p)\,W^2 - \bar p]$ in the r.h.s. of Eq.~(\ref{eq:HamCF3}) for each \ns{.} Superpose the terms if the binary involves a \dns{.}
	
	\item Solve the Hamiltonian constraint in the form given by Eq.~(\ref{eq:HamCF3}).
	
	\item If a \bh{,} compute the new irreducible $\hat M^\text{irr}_{1(2)}$, and if a \ns{} calculate the new rest mass $\hat M^0_{1(2)}$. Using $\xi_{1(2)}$ from Step 3, estimate the new ADM mass $\hat M_{1(2)}^\text{ADM} = \xi_{1(2)} \hat M^0_{1(2)}$. Notice that we are assuming that the ratio $\xi_{1(2)}$ does not change significantly from iteration to iteration.

	\item Next, identify the new mass $\hat M_{1(2)}$ with $\hat M_{1(2)}^\text{ADM}$  if a \ns{} and $\hat M_{1(2)}$ with $\hat M_{1(2)}^\text{irr}$ if a \bh{.} Calculate the new total mass $\hat M = \hat M_1+\hat M_2$ and mass ratio $\hat q = \hat M_1/\hat M_2$. If the new values differ from the values in Step 1 by more than a specified tolerance, adjust the bare masses of the \bh{} or central densities of the \ns{} according to a 2D secant algorithm \cite{kvaalen-1991}, and return to step 3.

\end{enumerate}

For the present work, we solve Eq.~(\ref{eq:HamCF3}) using a modified version of the
\twopunctures{} spectral code.  \twopunctures{} was originally developed by
Ansorg~\cite{Ansorg:2004ds} to construct \bbh{} initial data; that is, to solve Eq.~(\ref{eq:HamCF3}) with vanishing r.h.s. and $A_{ij}$ given by Eqs.~(\ref{eq:KP}) and/or (\ref{eq:KS}).

Once the conformal factor $\Phi$ is found from solving Eq.~(\ref{eq:HamCF3}), the spatial metric and extrinsic curvature are obtained from
$\gamma_{ij} = \Phi^4\eta_{ij}$ and
$K_{ij} =  \Phi^{-2}\bar A_{ij}$, respectively. The last step is constructing the hydrodynamical fields $\rho$, $p$, $W$ and $u^i$. Given $\Phi$, $\bar\rho_H$ and $\bar S^i$, we have that $\rho_H$ and $S^i$ are considered as known since $\rho_{\rm H}  = \Phi^{-8}\bar\rho_H$ and $S^i   = \Phi^{-10}\bar S^i$. On the other hand,
\begin{eqnarray}
\rho_{\rm H} &=& (\rho+p)\,W^2 - p\label{eq:uno} \\
S^i &=&  (\rho + p) W u^i \label{eq:tres}\,,
\end{eqnarray}
and from the second equation,
\begin{eqnarray}
\gamma_{ij}S^iS^j &=&  (\rho + p)^2 W^2\gamma_{ij} u^i u^j\nonumber\\
 &=&  (\rho + p)^2 W^2(W^2 -1)\label{eq:dos}\,,
\end{eqnarray}
where in the last equality we used that
 $ \gamma_{ij} u^iu^j = W^2 -1$ as implied by  $u^au_a = -1$. If we view that $p$ is given by an equation of state, Eqs.~(\ref{eq:uno}) and (\ref{eq:dos}) can be used to solve for $\rho$ and $W$. And the last step is to construct $u^i$ from Eq.~(\ref{eq:tres}).

\section{Tolman-Oppenheimer-Volkoff  model in isotropic coordinates}
\label{sec:TOV}

For the present work, we use a \tov{} stellar model to represent a \ns{}, with a polytropic equation of state $p= K\,\rho_0^\Gamma$ setting $\Gamma = 2$ and $K = 123.641\,M_\odot^2$. Since we assume conformal flatness, it is natural to recast the \tov{} model in isotropic coordinates. 
 \tov{} models are commonly constructed in coordinates in which the metric takes the form
\begin{equation}
\label{eq:tovs} 
ds^2 = -\alpha^2(\hat r)\,dt^2 + \left[1-\frac{2\,m(\hat r)}{\hat r}\right]^{-1}d\hat r^2 + \hat r^2\,d\Omega\,.
\end{equation}
On the other hand, the form of the metric (isotropic) compatible with our conformal flatness assumption is
\begin{equation}
 \label{eq:tovi}
ds^2 = -\alpha^2(r)\,dt^2 + \Phi(r)^4(dr^2 + r^2\,d\Omega)\,.
\end{equation}
In these coordinates, the equations that one needs to solve are the so called ``conformal thin sandwich'' equations~\cite{2010nure.book.....B}.
\begin{align}
&\partial^i \partial_i \Phi
= - \frac{1}{8}\Phi^{-7} \bar A_{ij} \bar A^{ij}  - 2\,\pi \Phi^5\,\rho_H \label{eq:TSham}\\
&\partial^j\partial_j \beta^i + \frac{1}{3} \partial^i\partial_j\beta^j = 2\,\bar A^{ij}\partial_j(\alpha\,\Phi^{-6}) + 16\,\pi\,\alpha\,\Phi^4\,S^i
\label{eq:TSshift}\\
&\partial^i\partial_i (\alpha\, \Phi) =
\alpha\,\Phi \left[\frac{7}{8} \Phi^{-8} \bar A_{ij} \bar A^{ij} + 2\,\pi \Phi^4(\rho_H+2\,S)\right]\label{eq:TSalpha}
\end{align}
where $\beta^i$ is the shift vector, $\rho_H$ is given by Eq.~(\ref{eq:rho_fluid}), $S^i$ by Eq.~(\ref{eq:si_fluid}) and $S = S^i\,_i$ with $S_{ij} = \gamma_i^{a}\gamma_j^{b}T_{ab}$.

For the metric (\ref{eq:tovi}), the conformal thin sandwich equations reduce to 
\begin{eqnarray}
\frac{1}{r^2}(r^2\,\Phi')' &=&  - 2\,\pi \Phi^5\,\rho \label{eq:TSham2}\\
\frac{1}{r^2}(r^2\,\Theta')' &=& 2\,\pi\,\Theta \Phi^4(\rho+6\,p)\label{eq:TSalpha2}
\end{eqnarray}
where primes denote differentiation with respect to $r$ and $\Theta \equiv \alpha\,\Phi$. Notice also that in this case $\beta^i = 0$, $A^{ij} = 0$, $S^i = 0$ and  $\rho_H = \rho$.  Finally, from $\nabla_bT^{ab} = 0$, one obtains
\begin{equation}
p' = -(\rho+p)\frac{\alpha'}{\alpha}= -(\rho+p)\left( \frac{\Theta'}{\Theta} - \frac{\Phi'}{\Phi} \right)\label{eq:P}
\end{equation}
Therefore, together with an equation of state, constructing \tov{} stellar models in isotropic coordinates involves solving Eqs.~(\ref{eq:TSham2}), (\ref{eq:TSalpha2}) and (\ref{eq:P}). Integration constants are chosen such that in the exterior of the star
\begin{eqnarray}
\Phi &=& 1+ \frac{M}{2\,r} \label{eq:extPhi}\\
\Theta &=& 1- \frac{M}{2\,r}\,,
\end{eqnarray}
with 
\begin{equation}
	\label{eq:mass}
	M = 2\pi \int_0^{r_0}  {r}^2 \Phi^{5}\rho \; dr
\end{equation}
the total mass of the star. Notice that $M = M_{\rm ADM}$ the ADM mass since Eq.~(\ref{eq:mass}) can be rewritten as Eq.~(\ref{eq:madm}).

If we denote by  $\Phi_{\rm tov}$, $\rho_{\rm tov}$ and $p_{\rm tov}$ the \tov{} solutions in isotropic coordinates, we then set
\begin{eqnarray}
\bar \rho &=& \Phi_{\rm tov}^8 \rho_{\rm tov}\\
\bar p &=& \Phi_{\rm tov}^8 p_{\rm tov}\,,
\end{eqnarray} 
and rewrite the Hamiltonian constraint Eq.~(\ref{eq:HamCF3}) as
\begin{eqnarray} 
&&\bar\Delta \Phi + \frac{1}{8}\Phi^{-7} \bar A_{ij} \bar A^{ij} = \nonumber\\
&&- 2\,\pi \Phi^{-3}\,\Phi^8_{\rm tov}[(\rho_{\rm tov}+p_{\rm tov})\,W^2 - p_{\rm tov}]    \label{eq:2punct}        
\end{eqnarray}
Notice that for an isolated \tov{} stellar model without linear or angular momentum ($\bar A_{ij} = 0$, $W = 1$ and $\Phi = \Phi_{\rm tov}$), Eq.~(\ref{eq:2punct}) reduces to  Eq.~(\ref{eq:TSham2}), namely
\begin{equation}
\bar\Delta \Phi_{\rm tov} = - 2\,\pi \Phi^5_{\rm tov}\,\rho_{\rm tov}\,.
\end{equation}

\section{Single Neutron Star with Linear Momentum}
\label{sec:single}

As a first test of the proposed method, we will consider an isolated \ns{} with linear momentum. We use a \tov{} stellar model with mass $M_* = 1.543 \,M_\odot$, radius $R_* = \unit[13.4]{km}$, and central density $\rho_c = 6.235\times 10^{14}\unit{gr\,cm^{-3}}$. We endow the star with linear momentum within the range $0 \le P/M_* \le 0.4$.

\begin{figure}
	\centering
	\includegraphics{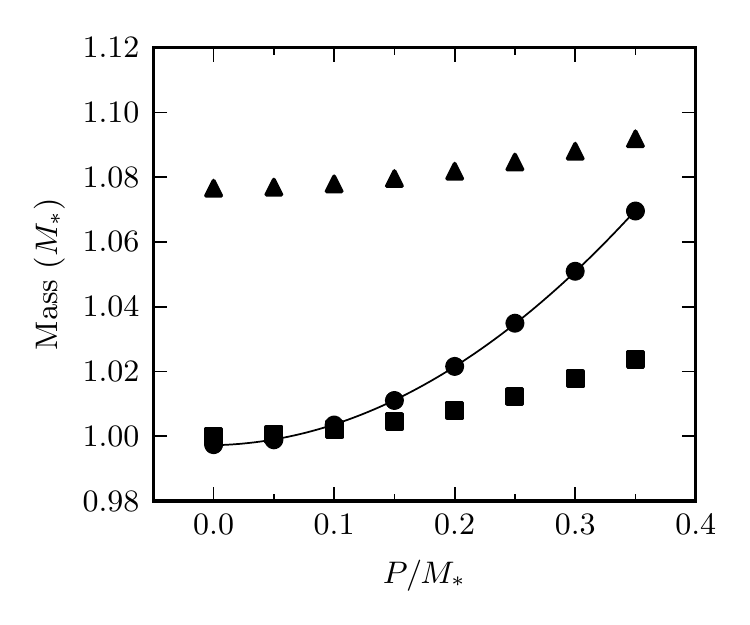}
\caption{ADM mass $M_{\rm ADM}$ (dots), rest mass $M_0$ (triangles) and $M_*W$ (squares) as a function of $P/M_*$ for a single \ns{}. Solid line represents a fit to $M_{\rm ADM} = M_* + cP^2$.}
\label{fig:ns_energy}
\end{figure}

Figure~\ref{fig:ns_energy} depicts with dots the ADM mass $M_{\rm ADM}$ as a function of $P/M_*$, and with triangles  the rest mass $M_0$. In the same figure, squares denote the quantity $M_*\,W$, where the Lorentz boost factor $W$ is calculated from~Eq.~(\ref{eq:boostP}). Notice that for small values of the linear momentum $M_{\rm ADM} \approx M_*\,W$. Also, it is not difficult to show from Eq.~(\ref{eq:madm}) and the Hamiltonian constraint (\ref{eq:2punct}) 
that $M_{\rm ADM} = M_* + O(P^2)$, consistent with the growth observed in Fig.~\ref{fig:ns_energy}.

\begin{figure}
	\centering
	\includegraphics{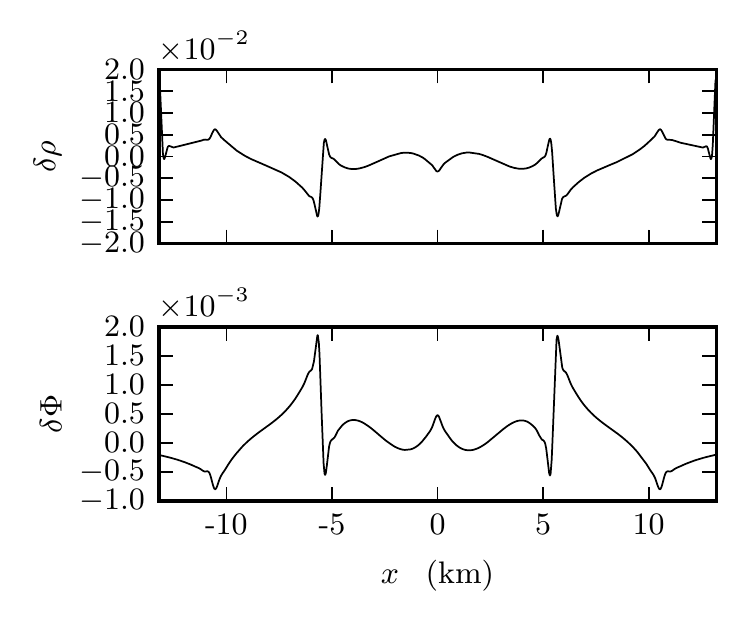}
	\caption{Relative differences along the $x$-axis between the \tov{} solution and the corresponding solution for a \tov{} star with momentum $P/M_* =  0.1$. Top panel shows the relative differences $\delta\rho$ in total mass-energy and bottom panel those in the conformal factor $\delta\Phi$.}
	\label{fig:ns_initial_rho}
\end{figure}

To further understand the changes that the momentum introduces to the \tov{} solution, we plot in Fig.~\ref{fig:ns_initial_rho} the relative differences with respect to the \tov{} solution of the total mass-energy density $\rho$ (top panel) and conformal factor $\Phi$ (bottom panel) along the $x$-axis, after solving the Hamiltonian constraint for a star with a linear momentum $P/M_* = 0.1$. The relative differences are computed as follows:
\begin{eqnarray}
\delta\rho &=& \frac{\rho-\rho_{\rm tov}}{\rho_{\rm tov}} \label{eq:delta}\\
\delta\Phi &=& \frac{\Phi-\Phi_{\rm tov}}{\Phi_{\rm tov}}
\end{eqnarray}
The differences in the mass-energy density are entirely due to the conformal factor. From $\rho = \Phi^{-8}\,\bar\rho$ and $\bar\rho = \Phi_{\rm tov}^8\,\rho_{\rm tov}$, one has that $\rho = ( \Phi/\Phi_{\rm tov})^{-8}\rho_{\rm tov}$, and thus from (\ref{eq:delta})
$
\delta\rho= (\Phi^{-8}-\Phi_{\rm tov}^{-8})/\Phi_{\rm tov}^{-8}\,.
$

\begin{figure}
	\centering
	\includegraphics{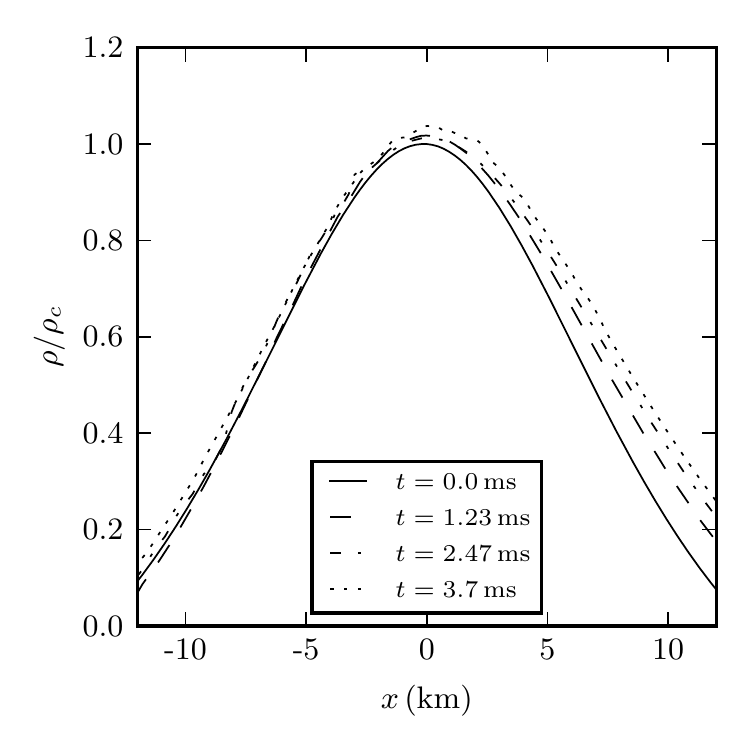}
	\caption{Density $\rho$ profiles along the $x$-axis for a \tov{} star with $P/ M_* = 0.1$ at various times throughout the evolution. The profiles have been normalized to the initial central density $\rho_c$ and shifted to be centered at $x=0$.  }
\label{fig:ns_evolution_rho}
\end{figure}

\begin{figure}
	\centering
	\includegraphics{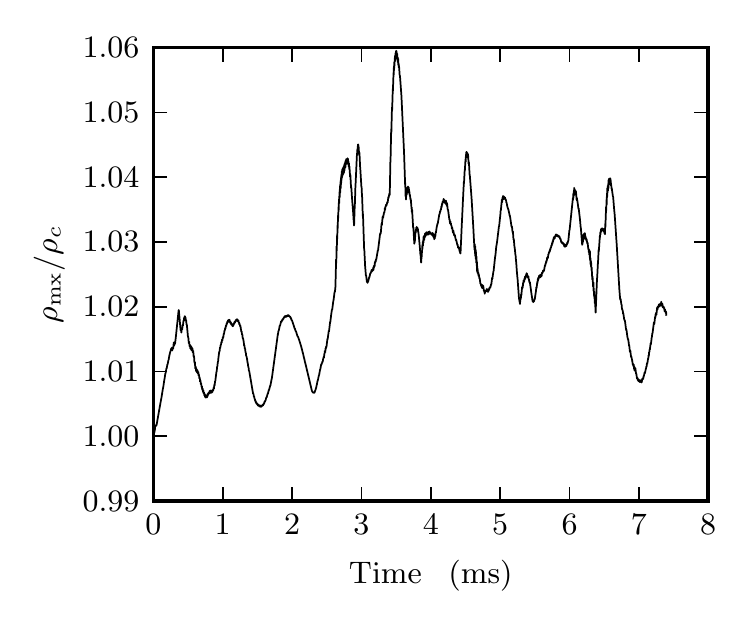}
\caption{Evolution of the central density  of the star in Fig.~\ref{fig:ns_evolution_rho} normalized to the initial central value $\rho_c$.}
\label{fig:ns_rho_maximum}
\end{figure}

In general terms, the evolutions of  the initial data for a single neutron star with linear momentum were satisfactory. The evolutions were carried out with the same gauge conditions used for puncture \bh{} evolutions~\cite{Campanelli:2005dd,Baker:2005vv}. We  noticed, however, few percent variations in the size and internal structure in the star during the course of the evolution. The changes in the size of the star are shown in Fig.~\ref{fig:ns_evolution_rho}, where we superimpose density profiles from different times for the case of a star with $P/M_* = 0.1$. Notice that the deformations are more prominent in the leading edge of the star (i.e. positive axis). Oscillations reveal themselves also in the central density of the star. Fig.~\ref{fig:ns_rho_maximum} shows the evolution of the central density in the star for the same case.

\section{Compact Object Binary Evolutions}
\label{sec:results}

Next, we test the performance of our prescription to construct initial data with evolutions of \dns{} and \bh{-}\ns{} binary systems.

\subsection{Non-spinning Double Neutron Star Binary}

\begin{figure*}
	\subfloat[Coordinate trajectory of one of the \ns{s}. \label{fig:bns_orbital_traj}]{
		\includegraphics[width=.45\textwidth]{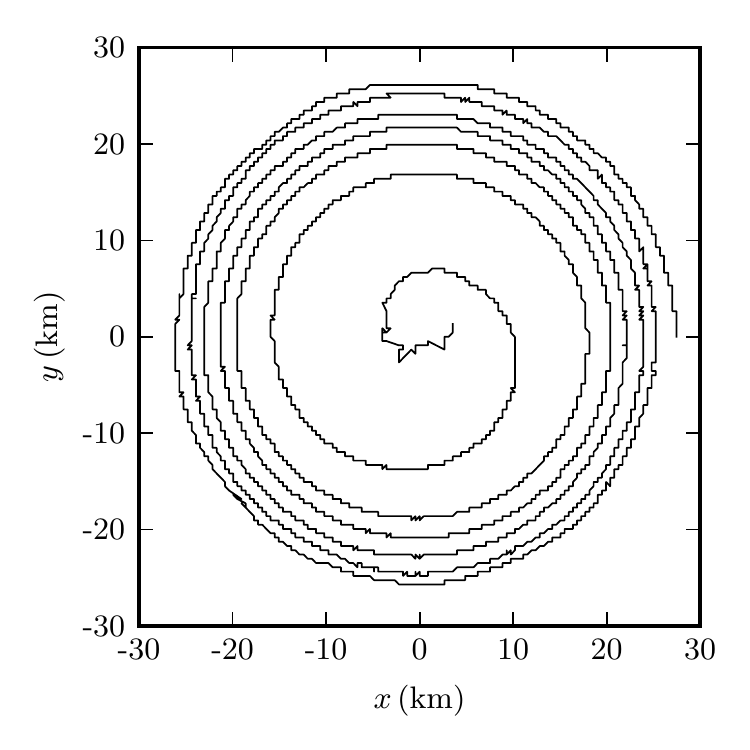}
	}
	\subfloat[Binary coordinate separation.  \label{fig:bns_orbital_sep}]{
		\includegraphics[width=.45\textwidth]{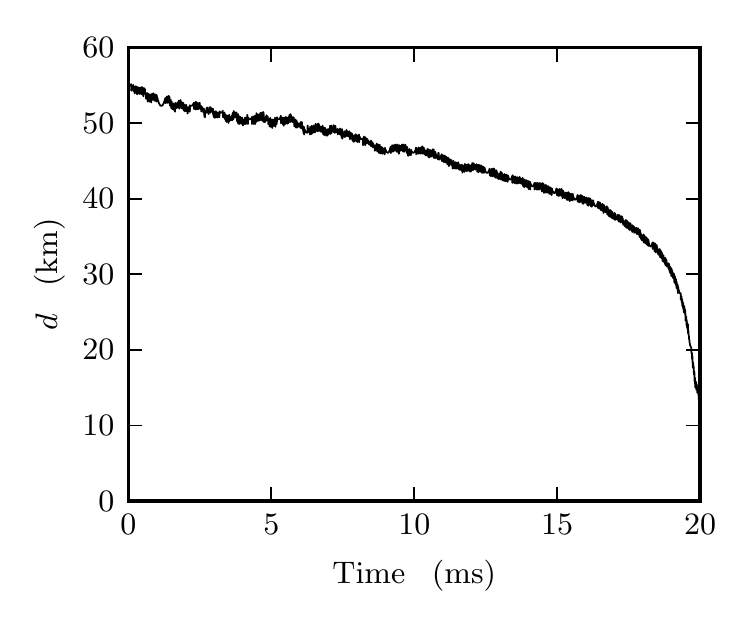}
	}

	\subfloat[Maximum rest mass density normalized to the initial central density $\rho_c$. \label{fig:bns_rhomax}]{
		\includegraphics[width=.45\textwidth]{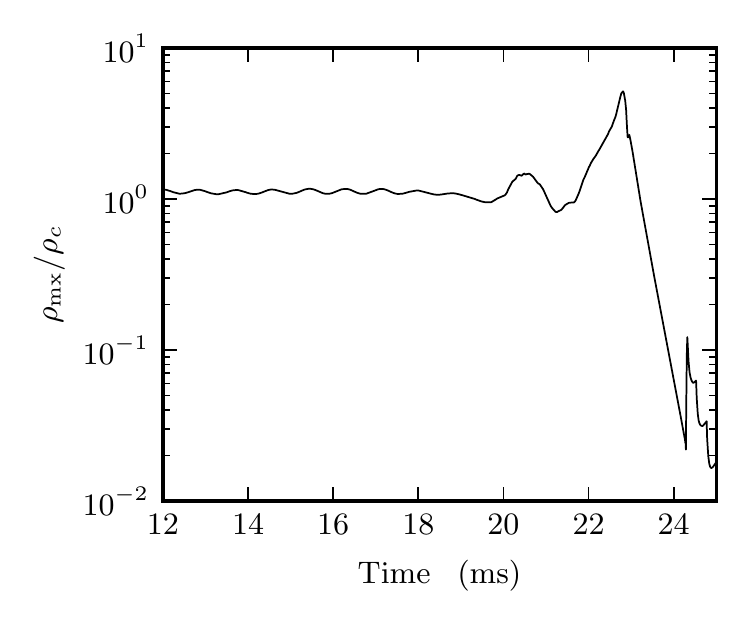}
	}
	\subfloat[Mode 2,2, of the Weyl scalar $\Psi_4$.\label{fig:bns_psi4_full}
	]{
		\includegraphics[width=.45\textwidth]{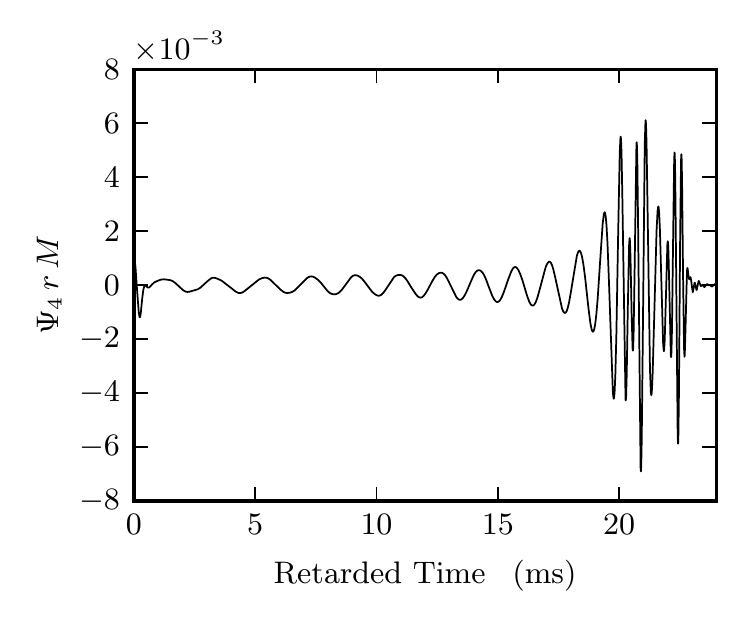}
	}
	\caption{Non-spinning \ns{} binary system.}
	\label{fig:bns_all}
\end{figure*}

We consider first an equal-mass \dns{} system. The \ns{s} have a  mass of $1.568\, M_\odot$, coordinate radius $13.1$ km, and they are initially separated by \unit[54.6]{km}. The configuration is similar to the model 1.62-45  in~\citet{baiotti-2008}. In their case, the stars have a mass of $1.62 \,M_\odot$, and their initial coordinate separation is \unit[45]{km}. 
The results of this simulation were obtained using 7 levels of mesh refinement. The finest mesh had resolution of $0.150\,M_\odot = 0.221$ km and extent of $26.6$ km. The wave-zone grid resolution was $9.58\,M_\odot = 14.1$ km.

Figure~\ref{fig:bns_orbital_traj} shows the coordinate trajectory of one of the \ns{} stars and Fig.~\ref{fig:bns_orbital_sep} the corresponding coordinate separation of the binary. The data in both figures end at the ``point-of-contact" (PoC), which occurs at approximately $\unit[18]{ms}$ after the start of the simulation or at a separation of approximately $25$ km. A \hmns{} forms $\unit[4]{ms}$ after the PoC, which collapses to a \bh{} in approximately $\unit[8]{ms}$. The collapse of the \hmns{} in~\citet{baiotti-2008} is $\unit[10]{ms}$, a difference that we attribute primarily to resolution effects.

\begin{figure*}
	\centering
	\includegraphics{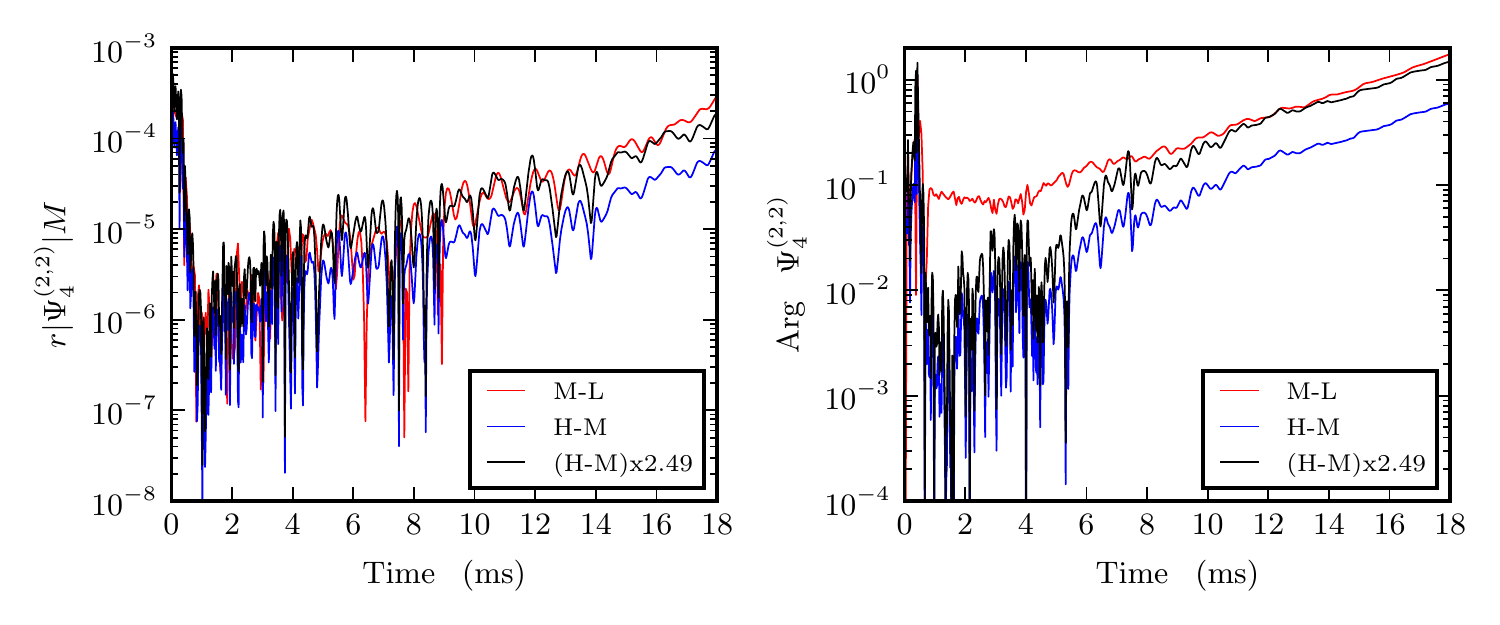}
	\caption{Amplitude (left panel) and phase (right panel) differences of the Weyl scalar $\Psi_4$ for three different resolutions of non-spinning \dns{} system simulations.  The resolutions in the finest grid are: $\unit[0.45]{km}$ (Low), $\unit[0.315]{km}$ (Medium), and $\unit[0.225]{km}$ (High). The (Medium--High) resolution is also presented in black re-scaled with a factor of 2.49, corresponding to 2\textsuperscript{nd} order convergence.}
	\label{fig:bns_convergence_order}
\end{figure*}

Figure~\ref{fig:bns_rhomax} shows the evolution of the central density normalized to its initial value. For comparison, see Fig.~12 in \citet{baiotti-2008}. The oscillations in Fig.~\ref{fig:bns_rhomax} for times earlier than $\unit[18]{ms}$ are similar, and likely due to the same reasons, to those seen in the case of a single \ns{} with linear momentum (see Fig.~\ref{fig:ns_rho_maximum}).  Since the amplitude of the oscillations decrease by increasing the initial separation of the binary, we suspect that the origin of the oscillations is because the \tov{} star has not been able to adjust to the linear momentum added and to the gravitational field by its companion. Similar oscillations have been observed in other initial data methods, for instance, in the work by~\citet{2013PhRvD..88f4060T}. We are currently investigating whether the prescription introduced by~\citet{2013PhRvD..88f4060T} to attenuate the oscillations will work in our case.

Figure~\ref{fig:bns_psi4_full} shows the 2,2 mode of the Weyl scalar $\Psi_4$, extracted at $462\,M_\odot$ from the binary, as a function of retarded time. At the beginning of the waveform, there is a small burst. This is the characteristic unphysical burst of radiation observed in \nr{} simulations that start with conformally flat initial data.  After the burst, $\Psi_4$ shows the expected chirp-like structure, the ringing of the HMNS during the time interval $\unit[18]{ms} \le t \le \unit[24]{ms}$, and the \qnm{} ring-down of the final \bh{.}  

\begin{figure*}
	\subfloat[{\unit[0]{ms}}]{\includegraphics[width=.45\textwidth]{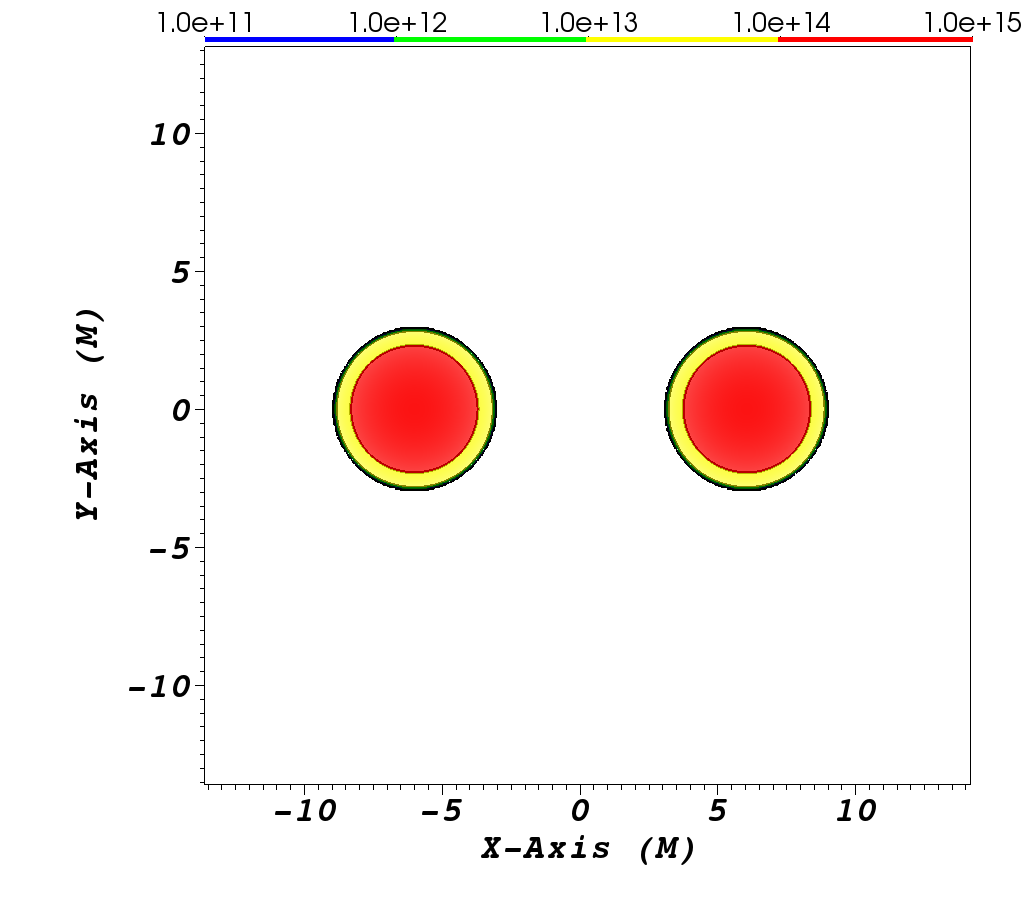}
}
	\subfloat[{\unit[20.3]{ms}}]{\includegraphics[width=.45\textwidth]{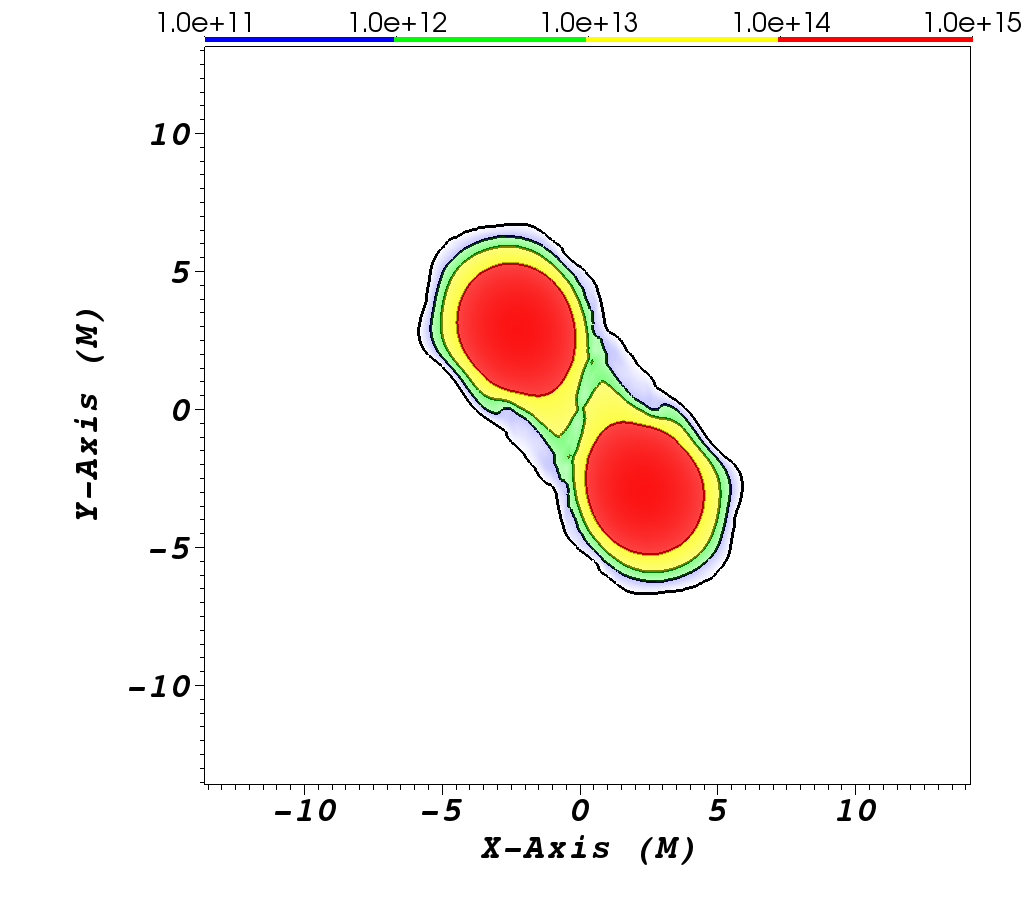}
}
		
	\subfloat[{\unit[26.9]{ms}}]{\includegraphics[width=.45\textwidth]{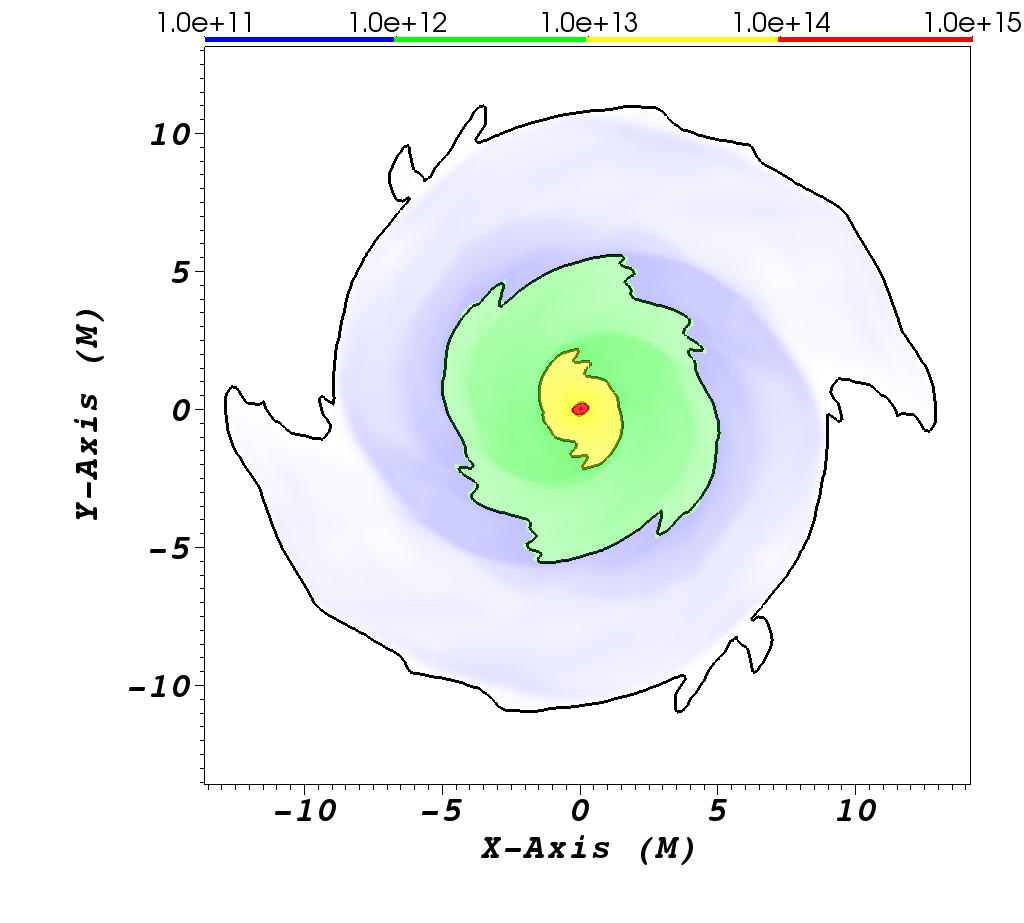}
}
	\subfloat[{\unit[24.6]{ms}}]{\includegraphics[width=.45\textwidth]{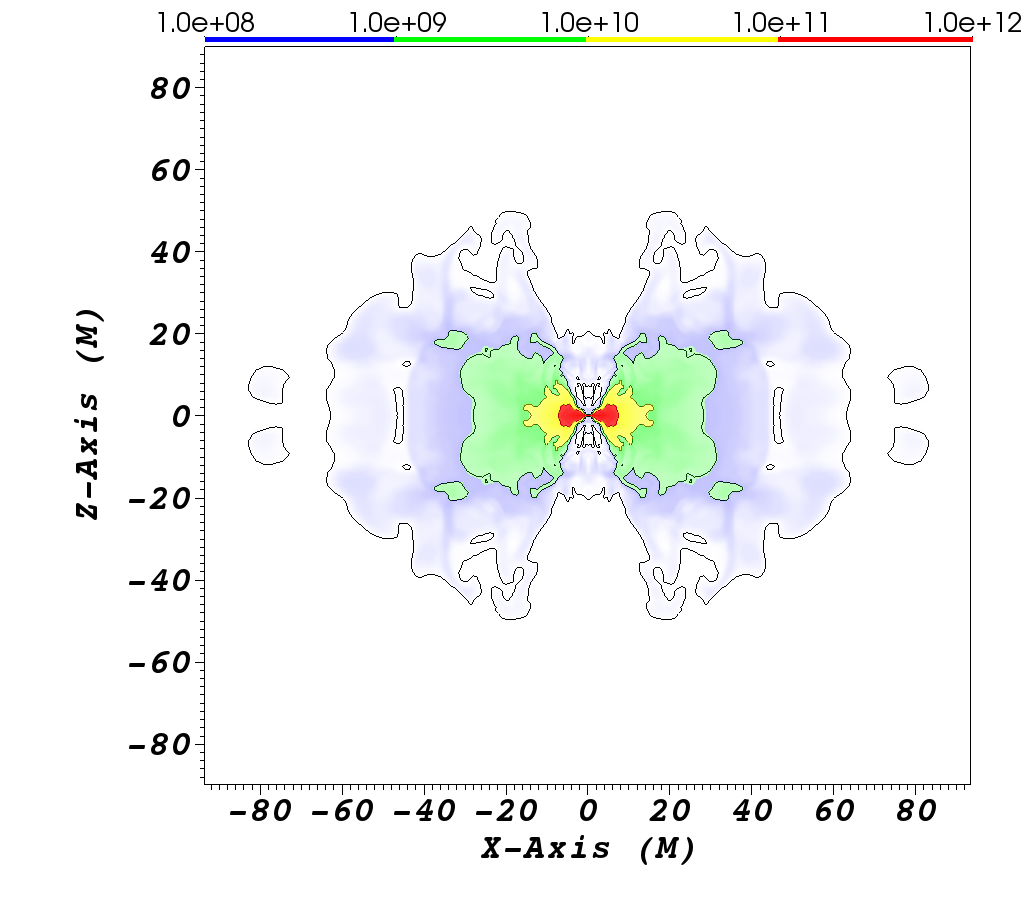}
}
	\caption{Rest-mass density snapshots from the non-spinning \dns{} binary evolution. Panels (a), (b) and (c) show the $xy$-plane and panel (d) the $xz$-plane   All densities are in units of $\unit{g \, cm^{-3}}$ and distances in units of $M = 3.14\,M_\odot$.}
	\label{fig:dns_slices}
\end{figure*}

Next, we analyze the convergence properties of the Weyl scalar $\Psi_4$, focusing only in the time segment before merger. We were unable to get ``clean" convergence estimates during the \hmns{} phase since numerical dissipation due to resolution effects leads to significant differences in the longevity of the resulting \hmns{}~\cite{hotokezaka-2013}. Figure~\ref{fig:bns_convergence_order} shows differences of amplitude and phase from three simulations with resolutions in the finest grid of $\unit[0.45]{km}$ (Low), $\unit[0.315]{km}$ (Medium), and $\unit[0.225]{km}$ (High). The red line shows the difference (Medium--Low) and the blue line (High--Medium). Assuming 2\textsuperscript{nd} order convergence, the three resolutions imply that  (Medium--Low) $\approx 2.49\times$(High--Medium). The black line in Fig.~\ref{fig:bns_convergence_order} depicts $2.49\times$(High--Medium) and thus consistency with 2\textsuperscript{nd} order convergence. For reference, the sector of the \maya{} code handling the geometrical fields is by design 6\textsuperscript{nd} order convergent. The hydrodynamical sector however is at best 3\textsuperscript{rd} order, but near shocks and local extrema can deteriorate to 1\textsuperscript{st} order, as seen in codes similar to ours where convergence order could be as low as 1.8~\cite{baiotti-2009}.

Finally, Fig.~\ref{fig:dns_slices} depicts snapshots of the rest-mass density during the evolution. Panels (a), (b) and (c) show the $xy$-plane and panel (d) the $xz$-plane. All densities are in units of $\unit{g \, cm^{-3}}$ and distances in units of $M = 3.14\,M_\odot$

\subsection{Spinning Double Neutron Star Binary}

\begin{figure*}
	\subfloat[Coordinate trajectory of one of the \ns{.} \label{fig:sbns_orbital_traj} ]{ \includegraphics[width=.45\textwidth]{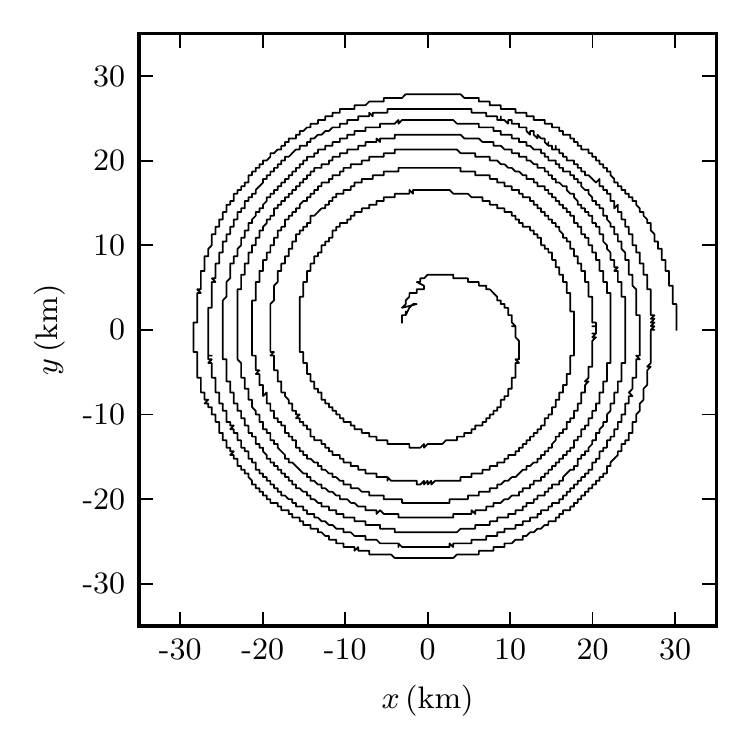} }
	\subfloat[Binary coordinate separation. \label{fig:sbns_orbital_sep} ]{\includegraphics[width=.45\textwidth]{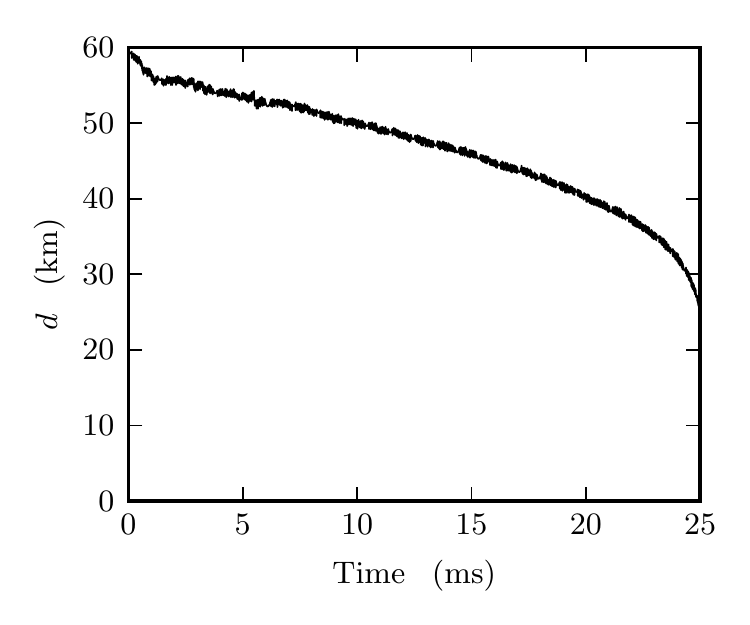}}

	\subfloat[Maximum rest mass density normalized to the initial central density $\rho_c$. \label{fig:sbns_rhomax} ]{ \includegraphics[width=.45\textwidth]{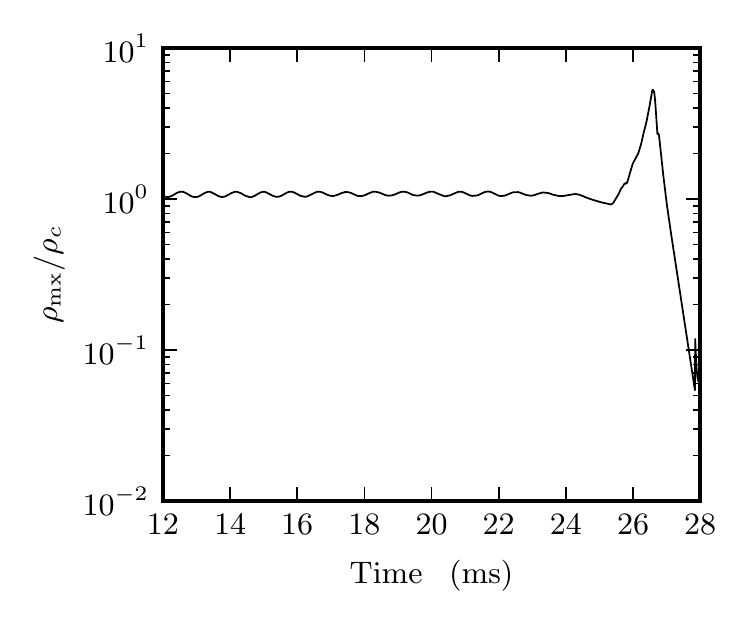} } 
	\subfloat[Mode 2,2, of the Weyl scalar $\Psi_4$.  \label{fig:sbns_psi4_full} ]{ \includegraphics[width=.45\textwidth]{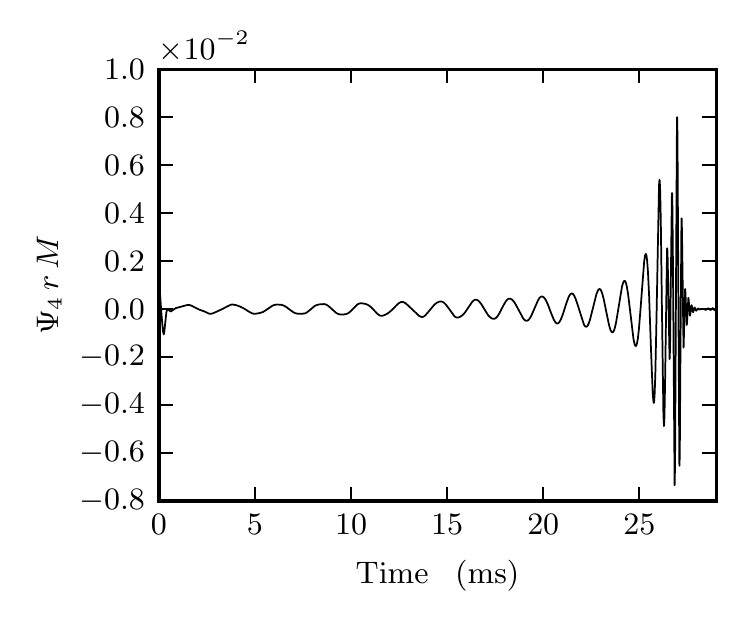} }

	\caption{Spinning \ns{} binary system}
	\label{fig:sbns_all}
\end{figure*}

The second example of evolution of initial data with the proposed scheme is again an equal-mass binary but now with spinning \ns{s}. Both stars have identical spins, anti-aligned to the orbital axis. The \ns{s} have a  mass of $1.57\, M_\odot$, coordinate radius $13.1$ km, and dimensionless spin parameter $\chi_s = -0.05$. At the beginning of the simulation, the \ns{s} are separated by \unit[61.2]{km}. With this choice of parameters, the binary system is similar to the case $\Gamma^{--}_{050}$ in \citet{bernuzzi-2014}. The grid structure is as follows:  the finest mesh has resolution $0.299 M_\odot = \unit[.442]{km}$ and extent $\unit[26.6]{km}$.  The radiation zone has resolution $19.2 M_\odot = \unit[28.3]{km}$.

Figure~\ref{fig:sbns_orbital_traj} shows the coordinate trajectory of one of the \ns{} stars and Fig.~\ref{fig:sbns_orbital_sep} the corresponding coordinate separation of the binary. Notice from Fig.~\ref{fig:sbns_orbital_traj} that the system performs 6 full orbits before merger.  Also noticeable is the slight kink or sudden drop in separation observed in Fig.~\ref{fig:sbns_orbital_sep} at the beginning of the evolution.  After the drop, the inspiral proceeds very smoothly, with minimal spurious eccentricity. As with the previous case, the data in both figures are depicted up to the PoC, which occurs at approximately $\unit[25]{ms}$ after the start of the simulation or at a separation of $\unit[26]{km}$. 

Figure~\ref{fig:sbns_rhomax} shows the evolution of the central density normalized to its initial value. Here again, we observe oscillations in the central density before merger. The \hmns{} forms at $\unit[26.2]{ms}$ and lasts for $\unit[1.3]{ms}$ before it collapses. From the waveform in Fig.~\ref{fig:sbns_psi4_full}, we notice that the \hmns{} undergoes two bursts. Also, the collapse to \bh{} is faster than in the non-spinning case. This is expected since the spins of \ns{} are anti-aligned with the orbital angular momentum and thus the \hmns{} is rotating slower than the \hmns{} in the non-spinning \dns{}.  The energy radiated is estimated to be approximately 0.7\% of total mass-energy, and the angular momentum radiated is 16\% of total angular momentum.  These values are slightly different form those reported by  \citet{bernuzzi-2014}---which are 1.2\% and 18\% respectively.

\begin{figure*}
		\subfloat[{\unit[0]{ms}}]{ \includegraphics[width=.45\textwidth]{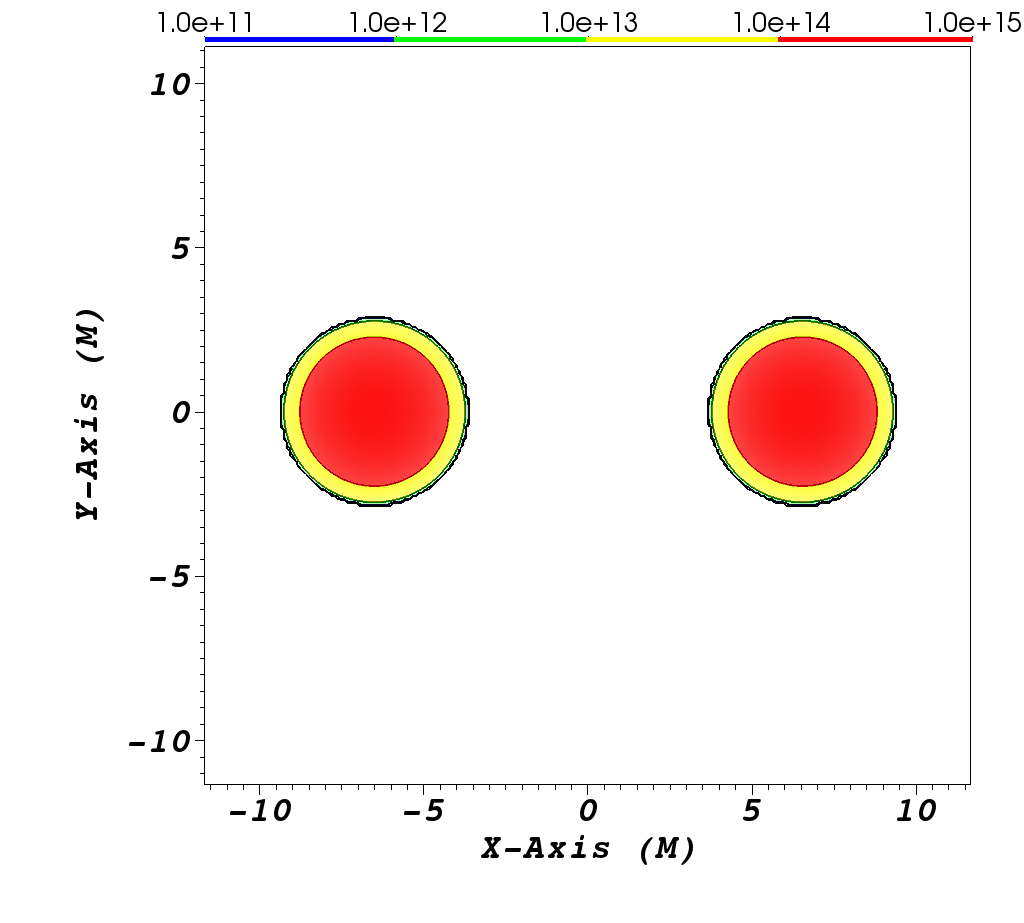}
}
		\subfloat[{\unit[24.6]{ms}}]{ \includegraphics[width=.45\textwidth]{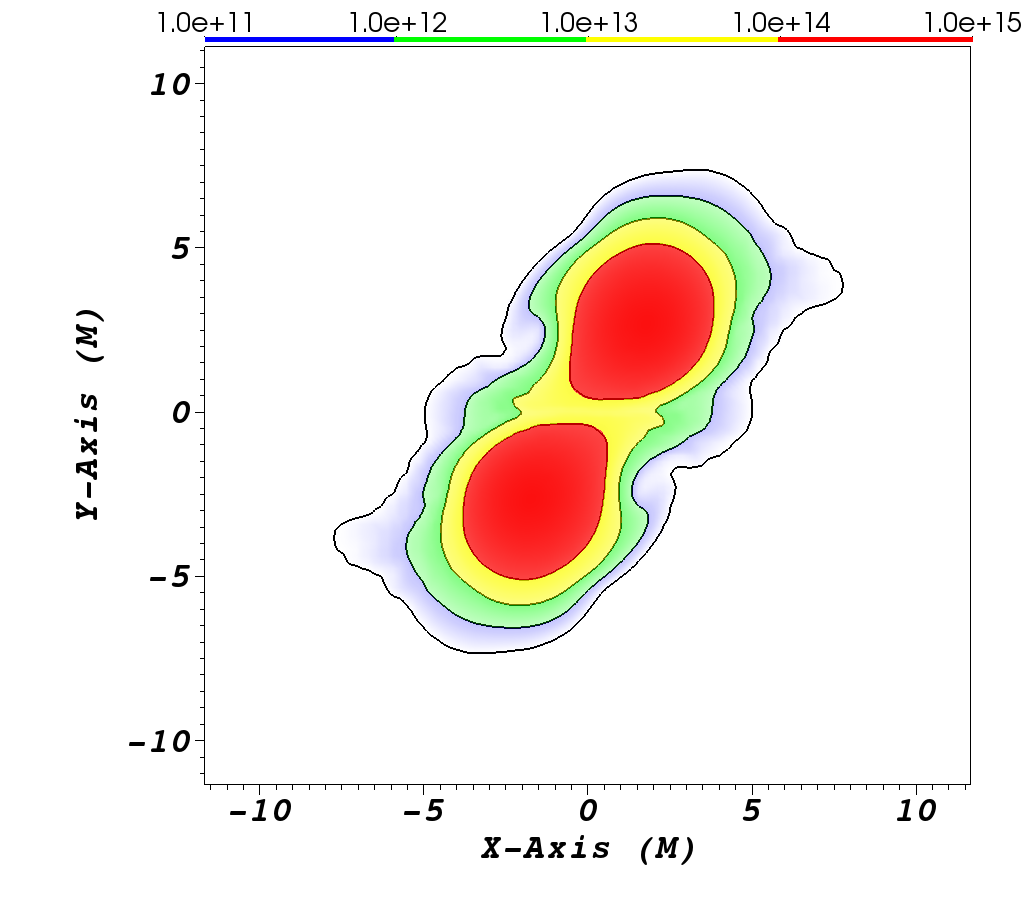}
}

		\subfloat[{\unit[26.1]{ms}}]{ \includegraphics[width=.45\textwidth]{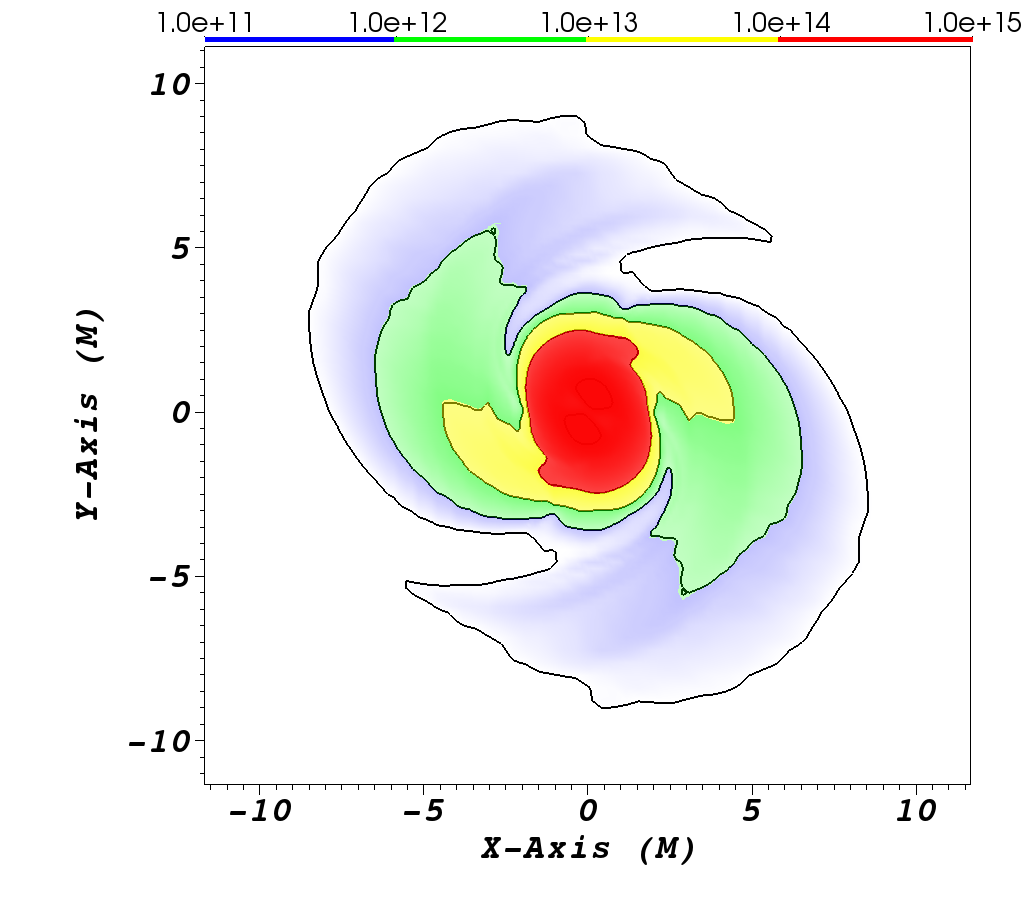}
}
		\subfloat[{\unit[29.0]{ms}} ]{ \includegraphics[width=.45\textwidth]{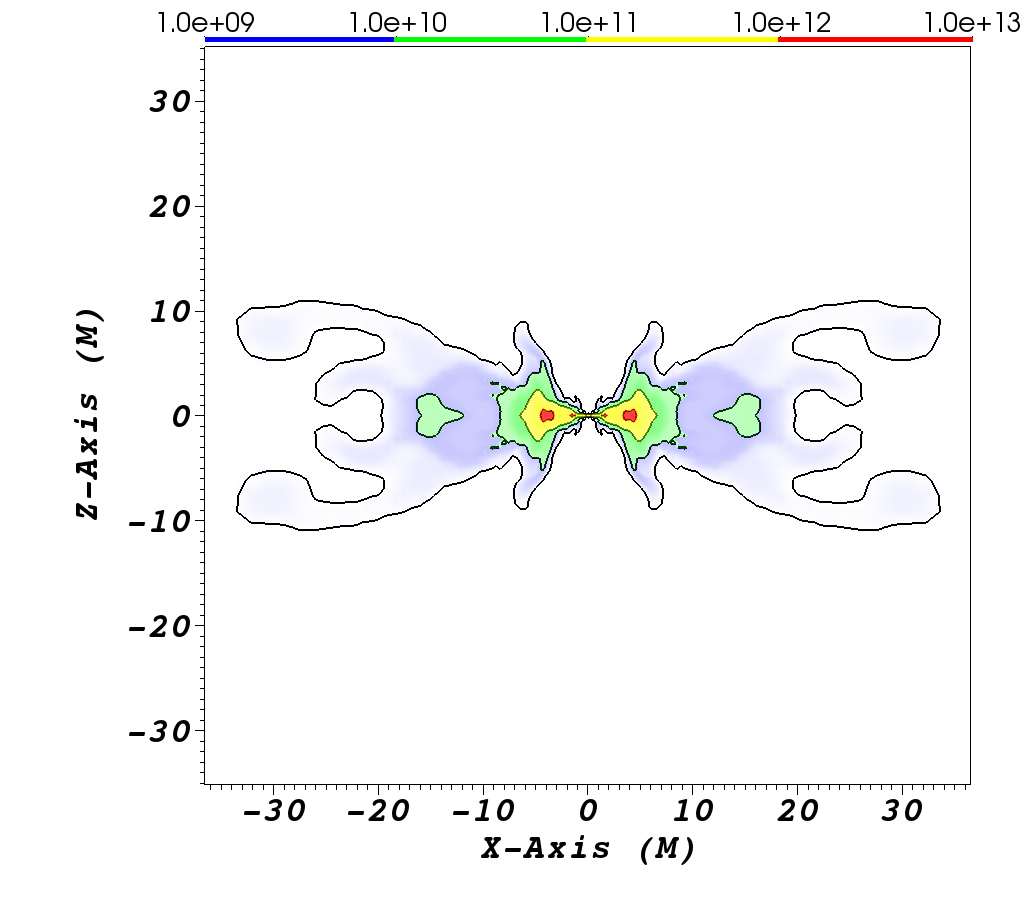}
}
	\caption{Rest-mass density snapshots from the spinning \dns{} binary evolution. Panels (a), (b) and (c) show the $xy$-plane and panel (d) the $xz$-plane   All densities are in units of $\unit{g \, cm^{-3}}$ and distances in units of $M = 3.14\,M_\odot$.}
	\label{fig:sdns_slices}
\end{figure*}

Finally, Fig.~\ref{fig:sdns_slices} depicts snapshots of the rest-mass density during the evolution. Panels (a), (b) and (c) show the $xy$-plane and panel (d) the $xz$-plane. All densities are in units of $\unit{g \, cm^{-3}}$ and distances in units of $M = 3.14\,M_\odot$.

\subsection{\bhns{} Binary}

\begin{figure*}
	\subfloat[Coordinate trajectories (\ns{} dashed and \bh{} solid)
		\label{fig:nsbh_orbital_traj}
	]{
		\includegraphics[width=.45\textwidth]{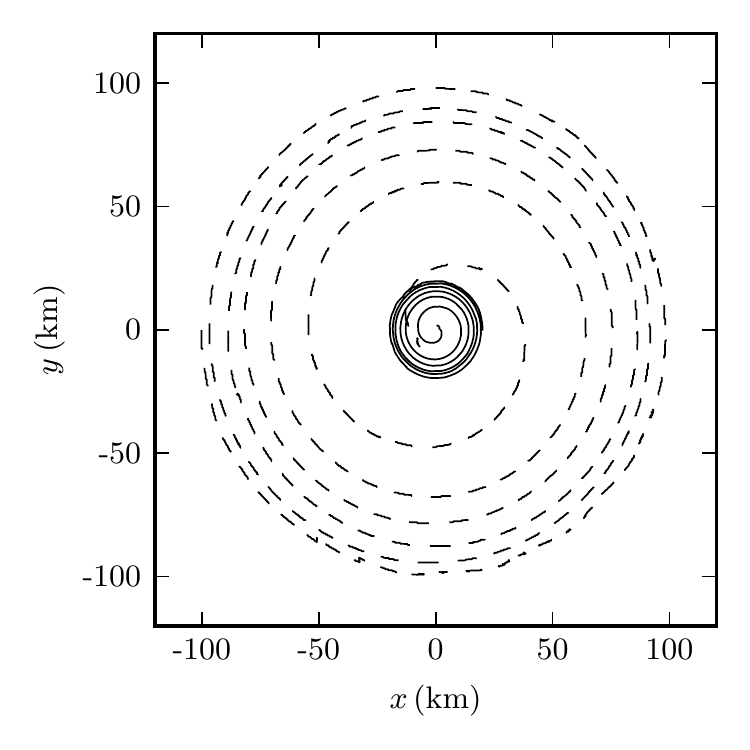}
	}
	\subfloat[Binary coordinate separation
		\label{fig:nsbh_orbital_sep}
	]{
		\includegraphics[width=.45\textwidth]{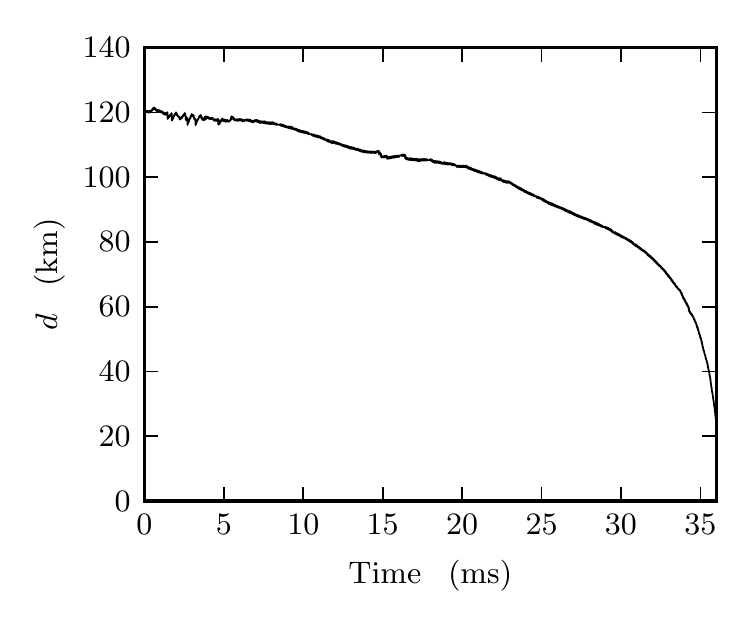}
	}

	\subfloat[Maximum rest mass density normalized to the initial central density $\rho_c$.
		\label{fig:nsbh_rhomax}
	]{
		\includegraphics[width=.45\textwidth]{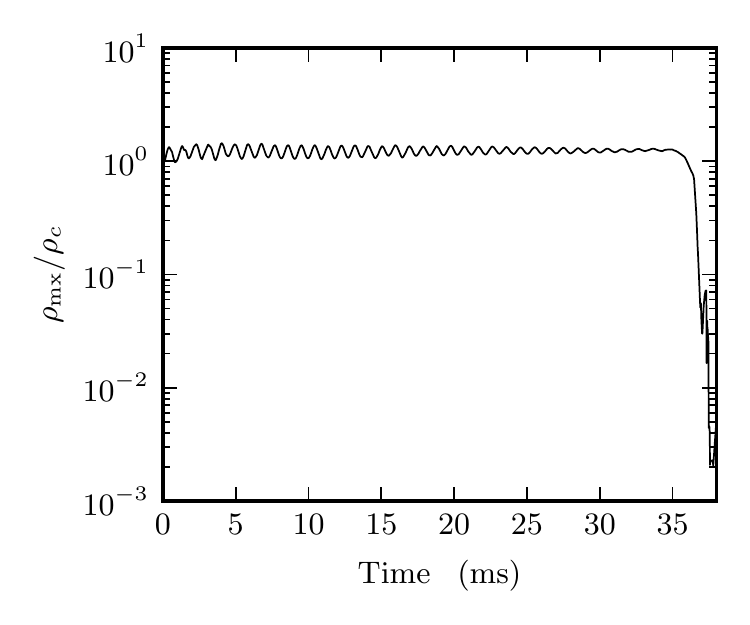}
	}
	\subfloat[Mode 2,2, of the Weyl scalar $\Psi_4$.
		\label{fig:nsbh_psi4_full}
	]{
		\includegraphics[width=.45\textwidth]{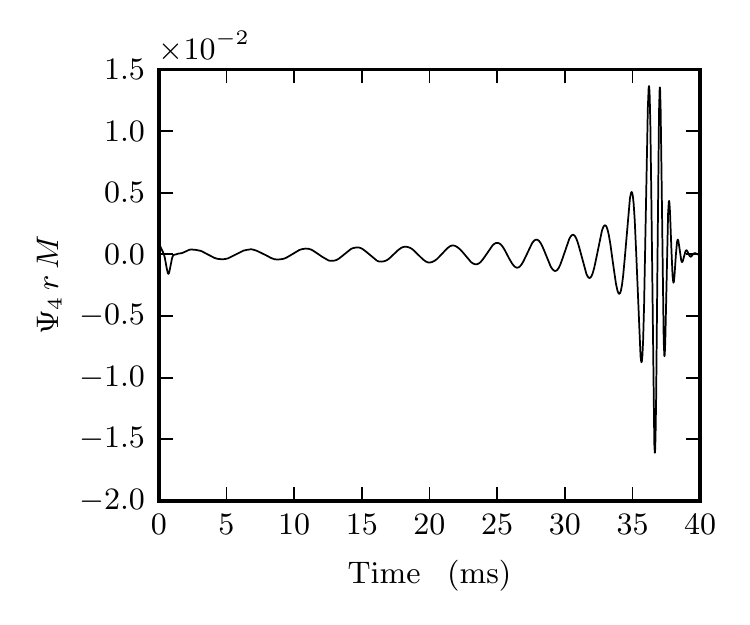}
	}
	\caption{\bhns{} binary system}
	\label{fig:nsbh_all}
\end{figure*}

The final example of evolution of initial data is for the case of  a \bhns{} binary system. The \ns{} has a mass of $1.54 M_\odot$ and a coordinate radius of $\unit[13.0]{km}$, and the \bh{} has a mass of $7.7\,M_\odot$ (i.e. 5:1 mass ratio binary). Both compact objects are non-spinning. The coordinate separation between the \bh{} and the \ns{} is $\unit[117]{km}$. With these parameters, the \bhns{} binary is similar to the M50.145b system in \citet{shibata-2009}. As with the \dns{} system, we cover the star with a single mesh whose side length is the diameter of the star. The grid structure has 8 levels of refinement, with finest resolution of $0.303 M_\odot = \unit[0.448]{km}$.  The finest mesh around the \bh{} has extent $9.10 M_\odot = \unit[13.4]{km}$.  The radiation zone has resolution of $38.8 M_\odot = \unit[57.3]{km}$.

Figure~\ref{fig:nsbh_orbital_traj} shows the trajectories of the \bh{} (solid line) and \ns{} (dashed line). The orbital separation of the binary is shown in Fig.~\ref{fig:nsbh_orbital_sep}. There is clear indication of spurious eccentricity. We attribute this eccentricity to the relatively small initial separation. Figure~\ref{fig:nsbh_rhomax} shows the maximum rest mass density during the course of the evolution.  The central density fluctuates as in the previous two cases, with the oscillations decaying at later times. The point at which the central density drops signals the time when the star is disrupted and swallowed by the \bh{}. This is also clear in the 2,2 mode of the Weyl scalar $\Psi_4$ (see Fig.~\ref{fig:nsbh_psi4_full}). At approximately $\unit[36]{ms}$, $\Psi_4$ shows the characteristic \qnm{} ringing of a \bh{.} 

\begin{figure*}[p]
		\subfloat[{\unit[0]{ms}}
]{
		\includegraphics[width=.45\textwidth]{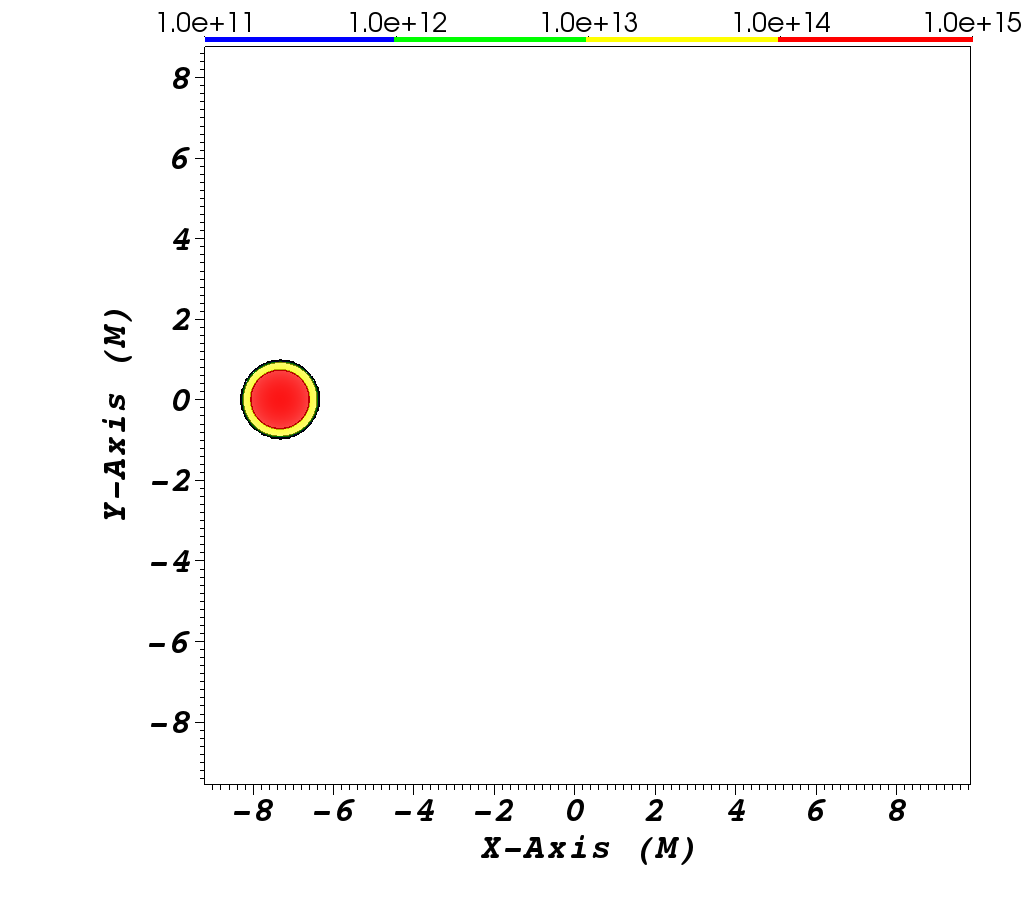}
}
		\subfloat[{\unit[34.4]{ms}}
]{
		\includegraphics[width=.45\textwidth]{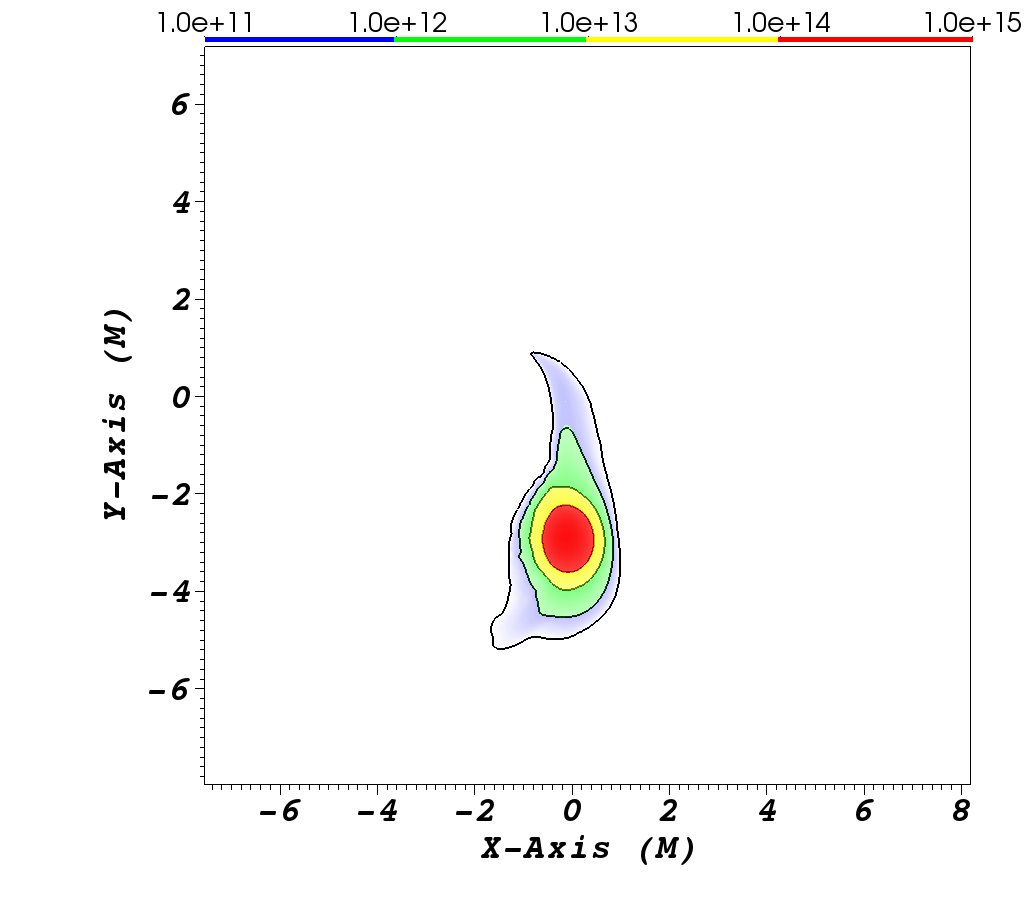}
}

		\subfloat[{\unit[35.9]{ms}}
]{
		\includegraphics[width=.45\textwidth]{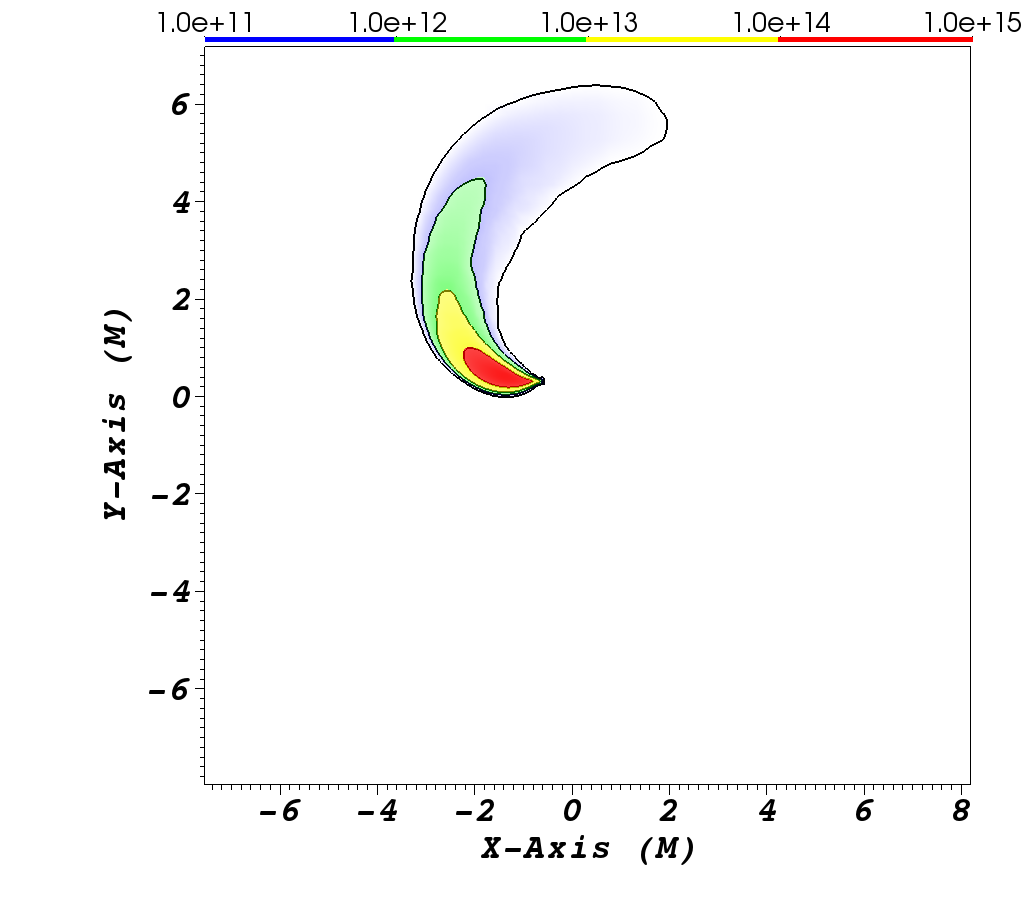}
}
		\subfloat[{\unit[37.9]{ms}}
]{
		\includegraphics[width=.45\textwidth]{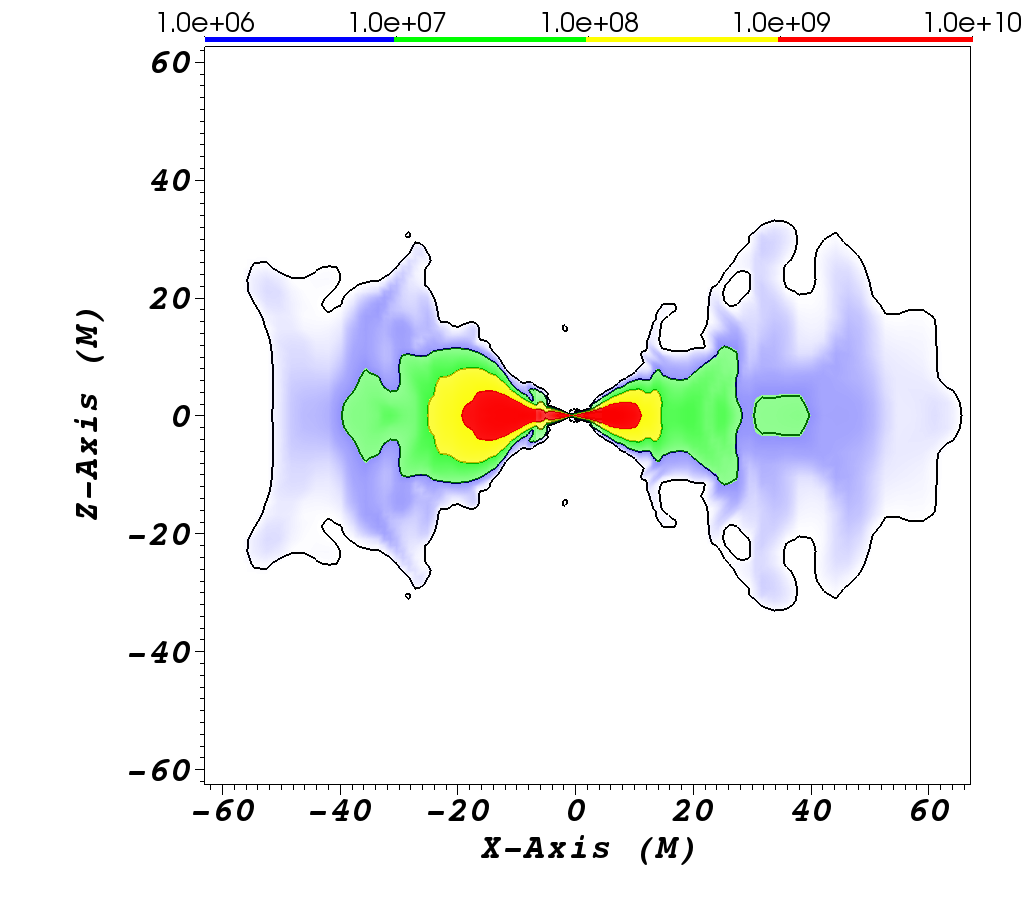}
}
	\caption{Rest-mass density snapshots from the bhns binary evolution. Panels (a), (b) and (c) show the $xy$-plane and panel (d) the $xz$-plane   All densities are in units of $\unit{g \, cm^{-3}}$ and distances in units of $M = 3.14\,M_\odot$.}
	\label{fig:nsbh_slices}
\end{figure*}

Figure~\ref{fig:nsbh_slices} depicts snapshots of the rest-mass density during the \bhns{} binary evolution. Panels (a), (b) and (c) show the $xy$-plane and panel (d) the $xz$-plane. All densities are in units of $\unit{g \, cm^{-3}}$ and distances in units of $M = 3.14\,M_\odot$

\section{Conclusions}
\label{sec:end}

We have introduced a new scheme to construct initial data for compact object binaries with \ns{} companions. The method is a generalization of the approach to construct initial data for \bbh{s} in which the \bh{s} are modeled as punctures and the extrinsic curvature is given by the Bowen-York solution to the momentum constraint~\cite{1980PhRvD..21.2047B}. In the method introduced in the present work, the extrinsic curvature for the \ns{s} is given by the solution derived by Bowen for spherically symmetric sources with linear momentum~\cite{1979GReGr..11..227B} and angular momentum~\cite{1989PThPh..81..360O}.
Given these extrinsic curvature solutions, we developed an iterative prescription to construct compact object binary initial data of \dns{s} or \bhns{s}. The prescription has a relatively low computational cost since it only requires solving the Hamiltonian constraint. As with the \bbh{} case, the method also allows one to specify the intrinsic and orbital parameters of the binary with direct input from \pnw{} approximations. The quality of the initial data method was demonstrated with a few examples of evolutions: an isolated \ns{} with linear momentum, \dns{} binaries, including spinning \ns{s}, and a \bhns{} system. The evolutions showed general agreement with similar cases found in the literature~\cite{baiotti-2008,bernuzzi-2014,shibata-2009}.

In this initial incarnation, the method was not devoid of defects.  The \ns{s} showed spurious breathing that translated into oscillations in their density structure. We are currently investigating applying the suggestion by~\citet{2013PhRvD..88f4060T} to mitigate the oscillations. In addition,  for \bhns{} binaries and \dns{} binaries with unequal masses, there is slight drift of the coordinate center-of-mass. In extreme cases, the drift  complicates waveform extraction.  
\acknowledgments

We thank P. Marronetti for helpful suggestions. This work was supported by NSF grants 1333360 and 1505824. Computations at XSEDE TG-PHY120016 and the Cygnus cluster at Georgia Tech.

\end{document}